\documentclass[superscriptaddress,amsmath,reprint,aps]{revtex4-2}
\usepackage[english]{babel}
\usepackage{graphicx, amsfonts, amssymb}
\usepackage{hyperref}
\usepackage{colortbl}
\usepackage{siunitx}

\usepackage[capitalize]{cleveref}
\crefname{section}{Sec.}{Sec.}

\usepackage[dvipsnames, svgnames, x11names]{xcolor}

\makeatletter
\newcommand*\dif{\mathop{}\!\mathrm{d}}
\newcommand{\rhos}{\rho_{s}}
\newcommand{\kb}{k}
\newcommand{\nn}{\nonumber\\}
\newcommand{\cit}[1]{Ref.~\cite{#1}}

\begin{document}
\title{On the time-dependent electrolyte Seebeck effect}
\author{Andr\'{e} Luiz Sehnem}\email{alsehnem@if.usp.br}
\affiliation{Institute of Physics, University of S\~{a}o Paulo, CEP 05508-090, S\~{a}o Paulo, Brazil}
\author{Mathijs Janssen}\email{mathijsj@uio.no}
\affiliation{{Department of Mathematics, Mechanics Division, University of Oslo, N-0851} Oslo, Norway}

\date{\today}
\begin{abstract}
Single-ion Soret coefficients $\alpha_{i}$ characterize the tendency of ions in an electrolyte solution to move in a thermal gradient.
When these coefficients differ between cations and anions, an electric field can be generated.
For this so-called electrolyte Seebeck effect to occur, the different thermodiffusive fluxes need to be blocked by boundaries---electrodes, for example.
Local charge neutrality is then broken in the Debye-length vicinity of the electrodes.
Confusingly, many authors point to these regions as the source of the thermoelectric field yet ignore them in derivations of the time-dependent Seebeck coefficient $S(t)$, giving a false impression that the electrolyte Seebeck effect is purely a bulk phenomenon.
Without enforcing local electroneutrality, we derive $S(t)$ generated by a binary electrolyte with arbitrary ionic valencies subject to a time-dependent thermal gradient. 
Next, we experimentally measure $S(t)$ for five acids, bases, and salts near titanium electrodes. 
For the steady state we find $S\approx\SI{2}{\milli\volt\per\kelvin}$ for many electrolytes, roughly one order of magnitude larger than predictions based on literature $\alpha_{i}$.
We fit our expression for $S(t)$ to the experimental data, treating the $\alpha_{i}$ as fit parameters, and also find larger-than-literature values, accordingly.
\end{abstract}
\maketitle

\section{Introduction}\label{sec:Introduction}
\begin{figure}
\includegraphics[width=7.4cm]{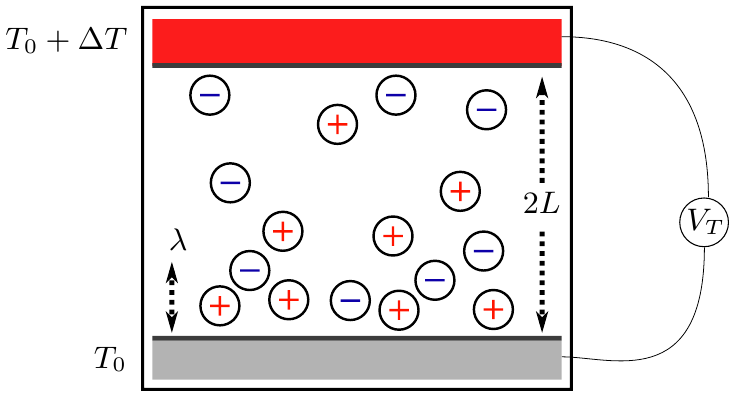}
\caption{Schematic (not to scale) of a thermoelectric cell comprising an electrolyte solution and two electrodes. 
Both anions and cations move to the cold electrode, the cations slightly more so than the anions. 
This gives a higher salt density near the cold electrode and a net charge density over a Debye length $\lambda$ near both electrodes. }
\label{fig:expsetup}
\end{figure}

Mobile charges often move preferentially along or against thermal gradients. 
For electrons in a metal, such movement underlies the Peltier-Seebeck effect, which is used in solid-state devices that refrigerate, measure temperature, or harvest thermal energy;
For ions in a nonisothermal fluid, thermodiffusion underlies the analogous electrolyte Seebeck effect \cite{Agar1963, Wurger2010}.
Microscopically, ionic thermodiffusion in electrolytes has been ascribed to ion-ion interactions \cite{helfand1960theory} and to the dynamics and structure of the surrounding fluid \cite{Eastman1926,*Eastman1928, Agar1963, tyrrell1967, agar1989, lin1973heats, Roemer2013, Sehnem2018}.
On mesoscopic length scales, ionic thermodiffusion can perturb the salt density $c(\mathbf{x})=\rho_{+}(\mathbf{x})+\rho_{-}(\mathbf{x})$---the Soret effect---with $\rho_{\pm}(\mathbf{x})$ being the local ionic densities of a binary electrolyte, leading to convection \cite{Caldwell1973, Gaeta1982} and variations in the electrolyte's conductivity \cite{Agar1960, *Agar1960a, Snowdon1960, *Snowdon1960a, Leaist1994} and refractive index \cite{Colombani1998, *Colombani1999, Roemer2013, Sehnem2018}. 
Ionic thermodiffusion can also perturb the local ionic charge number density $q(\mathbf{x})=z_{+}\rho_{+}(\mathbf{x})+z_{-}\rho_{-}(\mathbf{x})$, with $z_{\pm}$ being the ionic valencies. 
Resulting regions of nonvanishing ionic charge density $e q(\mathbf{x})$, with $e$ being the elementary charge, then generate a macroscopic thermoelectric field---the Seebeck effect---that can be measured as the thermovoltage $V_{T}$ between electrodes held at a temperature difference $\Delta T$ \cite{agar1957thermal, *Breck1957, Petit1987, Takeyama1988, Sanyal1988, Zhao2016, dilecce2018}; see \cref{fig:expsetup}. 
(The thermovoltage $V_{T}$ should not be confused with the thermal voltage $\kb T/e$, with $\kb$ being Boltzmann's constant and $T$ being temperature.)
The related Seebeck coefficient $S(t)=-V_{T}(t)/\Delta T(t)$ varies in time as ions take time responding to $\Delta T$, which itself may be time dependent. 

Eastman \cite{Eastman1926} and Wagner \cite{wagner1929} were the firsts to find expressions for the steady-state value $S_\mathrm{late}$ generated by a thermocell filled with a dilute multivalent, multicomponent electrolyte.
For a binary electrolyte near ideally polarizable electrodes \footnote{Conversely, when Faradaic currents are present, more terms enter $S_\mathrm{late}$ \cite{Agar1963, Bonetti2011, Bonetti2015, huang2015, Salez2017, salez2018, bhattacharya2020structural}.}, their expressions reduce to
\begin{equation}\label{eq:Seebecklate}
S_\mathrm{late}=\frac{2\kb}{e}\frac{\alpha_{+}-\alpha_{-}}{z_{+}-z_{-}}\,,
\end{equation}
with $\alpha_{i}=Q_{i}^{*}/(2\kb T)$ being single-ion Soret coefficient{s} ($i=+,-$), which is a modern dimensionless notation \cite{Wurger2010} for the ionic heat{s} of transport $Q_{i}^{*}$.
Eastman \cite{Eastman1928_2} and Wagner \cite{wagner1929} also found expressions for the Seebeck coefficient $S_\mathrm{early}$ generated instantaneously after applying a temperature difference, 
\begin{equation}\label{eq:Seebeckearly}
S_\mathrm{early}=\frac{2\kb}{e}\frac{D_{+}\alpha_{+}-D_{-}\alpha_{-}}{D_{+}z_{+}-D_{-}z_{-}}\,,
\end{equation}
with $D_{i}$ being the ionic diffusivities.

\Cref{eq:Seebecklate,eq:Seebeckearly} appeared frequently \cite{haasse1953, agar1957thermal, nakashima1992, Bonetti2015}, and so did simplifications for monovalent ions \cite{putnam2005, Vigolo2010, Wurger2010, majee2011, *majee2013, Eslahian2014, kim2018, dilecce2018} and extensions to multicomponent multivalent electrolytes \cite{de1947thermodynamique, Agar1963, Bonetti2011, Zinovyeva2014, huang2015, Salez2017, salez2018, bhattacharya2020structural}.
Derivations of \cref{eq:Seebecklate,eq:Seebeckearly} usually involve the ionic flux density, 
\begin{equation}\label{eq:flux}
\mathbf{J}_{i}(\mathbf{x},t)=-D_{i}\left(\mathbf{\nabla} \rho_{i}+\frac{z_{i}e\rho_{i}}{\kb T}\nabla \psi +\frac{\rho_{i}Q^{*}_{i}}{\kb T^2}\nabla T\right)\,,
\end{equation}
where, for notational ease, we have omitted the $(\mathbf{x},t)$-dependence of $T, \rho_{i}$, and the local electrostatic potential $\psi$, and of $D_{i}$ and $Q^{*}_{i}$, with indirect $(\mathbf{x},t)$-dependence through their $T$ and $\rho_{i}$-dependencies. 
The usual derivation of $S_\mathrm{late}$ asserts that breaking ionic charge neutrality comes at a huge energetic penalty \cite{newman2012electrochemical}, so that $q(\mathbf{x})=0$ must hold everywhere \cite{wagner1929,de1947thermodynamique, Agar1963, Vigolo2010,Wurger2010, Bonetti2011, huang2015, salez2018, kim2018, bhattacharya2020structural}. 
The steady-state expression $\mathbf{J}_{i}(\mathbf{x},t)=0$ then yields \cref{eq:Seebecklate}.
The usual derivation of $S_\mathrm{early}$ of an open circuit configuration [$\sum_{i}z_{i}\mathbf{J}_{i}(\mathbf{x})=0$] asserts that, at $t=0$, ions have not reacted to an applied thermal gradient yet, meaning that $\nabla \rho_{i}(\mathbf{x})=0$~\cite{de1947thermodynamique, Agar1963, huang2015, Salez2017, bhattacharya2020structural}\footnote{\cit{han2020} claims \cref{eq:Seebeckearly} to hold at late instead of at early times.}. 

As pointed out by Chikina \textit{et al.} \cite{chikina2012} the assumption $q(\mathbf{x})=0$ is incompatible with $S_\mathrm{late}\neq0$:
If ionic densities would be strictly unperturbed, the Poisson equation [cf. \cref{eq:Poisson}] would predict a spatially constant electrostatic potential, hence $S_\mathrm{late}=0$.
To consistently account for the ionic electrostatic interactions, it is imperative to solve both \cref{eq:flux} \textit{and} the Poisson equation \cite{chikina2012,majee2011}.
Doing so, one can find a steady-state solution for $\psi(\mathbf{x})$ [\cref{eq:psi_sol} of the supplementary material] that correctly yields \cref{eq:Seebecklate}.
Moreover, this $\psi(\mathbf{x})$ shows that charge neutrality is broken near the electrodes over a length $\lambda$ called the Debye length. 
While, indeed, breaking ionic charge neutrality is energetically costly, this does not forbid the system to break it in tiny---$\lambda\sim\SI{1}{\nano\meter}$ for aqueous electrolytes---regions close to the electrode surfaces, anyway \cite{newman2012electrochemical} \footnote{Even though local charge neutrality is broken near its boundaries, the system is still \textit{globally} charge neutral, $\int\dif\mathbf{x}\, q(\mathbf{x})=0$}. 
Ignoring those regions, however, misses the point that the thermovoltage is caused by ionic charge separation \cite{Note1}.

The above derivation of $S_\mathrm{early}$ suffers from the same ailment: unperturbed ionic densities $\nabla \rho_{i}(\mathbf{x})=0$ are incompatible with $S_\mathrm{early}\neq0$.
It is also unclear on what timescale the ``instantaneous" Seebeck coefficient $S_\mathrm{early}$ is reached. A thermal gradient cannot be imposed instantaneously, and even if it could be, ions would not react to it instantaneously. 
Moreover, $S(t)=0$ as long as ions have not reacted. 
The first step to resolving these problems was set by Stout and Khair \cite{StoutKhair2017}, who found the time-dependent thermovoltage $V_{T}(t)$ of a $z:z$, $D_{+}=D_{-}=D$ electrolyte, again, using both \cref{eq:flux} \textit{and} the Poisson equation.
Assuming an instantaneous steady-state temperature profile, they found that ${S}(t)$ relaxes to the $z:z$ simplification of \cref{eq:Seebecklate} on the Debye timescale $\tau_{D}=\lambda^2/D$.
As $\tau_{D}\sim\SI{1}{\nano\second}$ for typical aqueous electrolytes, Stout and Khair's finding suggested that the electrolyte Seebeck effect is much faster than the Soret effect, which develops on the diffusion timescale $\tau_\mathrm{dif}= 4L^2/(\pi^2 D)\sim \SI{e3}{\second}$ \cite{Agar1960}, with $L=\SI{3}{\milli\meter}$ being a typical value for the electrode separation $2L$.
This theoretical finding was at odds, however, with experiments that found that $V_{T}(t)$, indeed, develops fast at first [faster than the experimental resolution (seconds)], but then evolves with the slow timescale $\tau_\mathrm{dif}$ \cite{Bonetti2015, Salez2017}. 
Many authors, therefore, explained the dynamics of $V_{T}(t)$ in terms of the time-dependent salt density $c(\mathbf{x}, t)$ \cite{de1947thermodynamique, agar1957thermal, blokhra1972, lin1973heats, Petit1987, Bonetti2015, Chikina2015, gunnarshaug2019reversible}, shown by Bierlein \cite{Bierlein1955} to evolve with $\tau_\mathrm{dif}$ (see also Refs.~\cite{de1942theorie,agar1960rate}).
Yet, while the Poisson equation connects $q(\mathbf{x}, t)$ to $V_{T}(t)$, there is no such connection between $c(\mathbf{x}, t)$ and $V_{T}(t)$, which means that Bierlein's expression cannot be used to interpret $V_{T}(t)$.
To solve this problem, one of us---with Bier---generalized Stout and Khair's model to an electrolyte for which $D_{+}\neq D_{-}$ \cite{Janssen2019}. 
For this case, we showed that $q(\mathbf{x}, t)$ relaxes exponentially with both $\tau_{D}$ \textit{and} $\tau_\mathrm{dif}$---and, because of Poisson's equation, so does $V_{T}(t)$ [see Fig.~5(b) therein].
Our analytical expression for $V_{T}(t)$ yielded $S=0$ strictly at $t=0$---respecting the initial conditions inserted in Poisson's equation---and also took an elegant form (cf.~Eq.~17 of \cit{Janssen2019}) at \textit{intermediate} times for which $t\gg \tau_{D}$ and $t \ll \tau_\mathrm{dif}$. 
Strikingly, Eq.~(17) of \cit{Janssen2019} coincides with the $1:1$ simplification of \cref{eq:Seebeckearly}---providing evidence that the ``instantaneous" Seebeck coefficient is rather reached on the Debye timescale---though we did not realize this at the time of writing \cit{Janssen2019}.

Here, we generalize the $S(t)$ {derivation} of \cit{Janssen2019} for a $1:1$ electrolyte subject to an instantaneous thermal gradient to a {binary electrolyte with arbitrary ionic valencies} subject to a temperature difference $\Delta T(t)$ that increases with a characteristic timescale $\tau_\mathrm{ap}$. 
We confirm \cref{eq:Seebecklate} for $t\gg \tau_\mathrm{ap},\tau_\mathrm{dif}$ and \cref{eq:Seebeckearly} for $t\ll\tau_\mathrm{ap},\tau_\mathrm{dif}$. 
Unlike earlier derivations of \cref{eq:Seebecklate,eq:Seebeckearly}, ours does not enforce local charge neutrality.
To test our theory, in \cref{sec:experiments}, we determine $S(t)$ for five aqueous binary electrolytes containing monovalent and divalent ions (hydroxides, acids, and chloride and sulfate salts) near Ti electrodes.
Our theory and experiments agree qualitatively on $S(t)$ transitioning from $S_{\rm early}$ to $S_{\rm late}$ plateaus, and even decently predicts the transition time between those plateaus.
The height of the $S(t)$ plateaus, however, does not compare well with predictions from \cref{eq:Seebecklate,eq:Seebeckearly} using literature $\alpha_{i}$. 
We speculate about possible causes of this discrepancy in \cref{sec:Discussion}.

\section{Theory for the time-dependent electrolyte Seebeck coefficient $S(t)$}\label{sec:theory}
We consider a {binary electrolyte with arbitrary ionic valencies $z_{+}$ and $z_{-}$, diffusion constants $D_{+}$ and $D_{-}$, } and salt concentration $\rhos$ between two flat parallel ideally-polarizable electrodes. 
We choose the electrode separation $2L$ to be much smaller than the electrode's size in the lateral direction. 
Moreover, we assume the system to be isothermal in this lateral direction.
Under these conditions, all observables depend only on the coordinate $x\in[-L,L]$.
We treat the solvent as a structureless dielectric medium of dielectric constant $\varepsilon(x,t)$, which could vary spatiotemporally through $c(x,t)$ and $T(x,t)$-dependencies. 

We subject this setup to a time-dependent temperature difference $\Delta T(t)$ that we choose as follows. 
The temperature difference in the experiments below increases roughly as $\Delta T(t)=\Delta T_{\infty} \left[1-\exp (-t/\tau_\mathrm{ap})\right]$, with $\Delta T_{\infty}$ its late-time value and with $\tau_\mathrm{ap}=\SI{43}{\second}$ being a characteristic timescale \cite{Note4}.
Theoretically, when the temperature of one of the electrodes would increase stepwise, the electrolyte would relax thermally with the timescale $\tau_{T}=4L^{2}/(\pi^2 a)$, with $a$ being the thermal diffusivity \cite{carslaw1959conduction}. 
Our experiments on aquaeous electrolytes are characterized by $2L=\SI{5}{\milli\meter}$ and $a=\SI{1.4e-7}{\meter\squared\per\second}$, yielding $\tau_{T}=\SI{18}{\second}$.
As $\tau_\mathrm{ap}\gtrapprox \tau_{T}$ for our experiments, in our theoretical model we consider a simple form of $T(x,t)$ wherein $\Delta T(t)$ builds up exponentially, but where $T(x,t)$ is quasi-equilibrated to a profile linear in $x$: $T(x,t)=T_{0}+\Delta T_{\infty} \left[1-\exp (-t/\tau_\mathrm{ap})\right] (x+L)/(2L)$.
We expect the error in $V_{T}(t)$ that we make by not solving the heat equation to be small:
Reference \cite{Janssen2019} theoretically studied the response on electrolyte to a suddenly applied potential---that is, applied much faster than our temperature difference $\Delta T(t)$.
They found that an analytical approximation to $V_{T}(t)$ derived under the assumption that $T(x,t)$ reached its steady state instantaneously---a more radical assumption than made here---accurately reproduced numerical calculations for $V_{T}(t)$ wherein the spatiotemporal variation of $T(x,t)$ was fully accounted for.

For the above $T(x,t)$, \cref{eq:flux} simplifies to 
\begin{equation}\label{eq:1dflux}
-\frac{J_{i}}{D_{i}}=\partial_{x} \rho_{i}+\frac{z_{i}e\rho_{i}}{\kb T}\partial_{x} \psi +\frac{\rho_{i}Q^{*}_{i}}{\kb T^2}\frac{\Delta T_{\infty } \left[1-\exp (-t/\tau_\mathrm{ap})\right]}{2L}\,.
\end{equation}
With the Poisson equation and the continuity equation
\begin{subequations}\label{eq:electrokinetic}
\begin{align}
\varepsilon_{0}\varepsilon \partial^{2}_{x} \psi&=- e q \,,\label{eq:Poisson}\\
\partial_{t} \rho_{i}&=-\partial_{x} J_{i}\,,\label{eq:continuity_ions}
\end{align}
\end{subequations}
with $\varepsilon_{0}$ being the vacuum permittivity, we come to a closed set of equations for $\rho_{\pm}(x,t)$ and $\psi(x,t)$.
Again, for readability we have dropped the $(x,t)$-dependence of all observables and parameters. 

We subject \cref{eq:1dflux,eq:electrokinetic} to the following initial and boundary conditions
\begin{subequations}\label{eq:initandbc}
\begin{align}
\rho_{i}(x,t=0)&=\rho_{i,0}\,,\label{eq:initandbca} \\
q(x,t=0)&=\rho_{+,0}z_{+}+\rho_{-,0}z_{-}=0\,,\label{eq:initandbcb}\\
\partial_{x}\psi(\pm L,t)&=0\,,\label{eq:initandbcc} \\
J_{i}(\pm L,t)&=0\,,\label{eq:initandbcd}
\end{align}
\end{subequations}
which express initially homogenous ionic density profiles [\cref{eq:initandbca}] of a system with local (and thus also global) charge-neutrality at $t=0$ [\cref{eq:initandbcb}].
We consider a thermoelectric cell in open-circuit configuration and assume its electrodes to be ideally polarizable and nonadsorbing. 
Hence, the electrodes remain uncharged, expressed by Gauss's law in \cref{eq:initandbcc}, and ionic currents at the electrodes vanish [\cref{eq:initandbcd}].
Notably, \cref{eq:initandbcc} only fixes $\psi$ up to a constant irrelevant to the thermovoltage $V_{T}(t)=\psi(L,t)-\psi(-L,t)$, the quantity of interest here.
This definition of $V_{T}$ is the same as W\"{u}rger's \cite{Wurger2010,wurger2020thermopower} and differs from \cit{Janssen2019} by an overall minus sign. 
We see from \cref{eq:Poisson,eq:initandbcb} that $\psi(x,t=0)=cst.$, giving $V_{T}(t=0)=0$ and $S(t=0)=0$.
While the diffusion ($\sim \partial_{x}\rho_{i}$) and electromigration term ($\sim \partial_{x}\psi$) of \cref{eq:1dflux} vanish at $t=0$, the finite thermodiffusion term ($\sim \Delta T_{\infty}$) drives the system out of equilibrium.
\begin{figure}
\includegraphics[width=8.6cm]{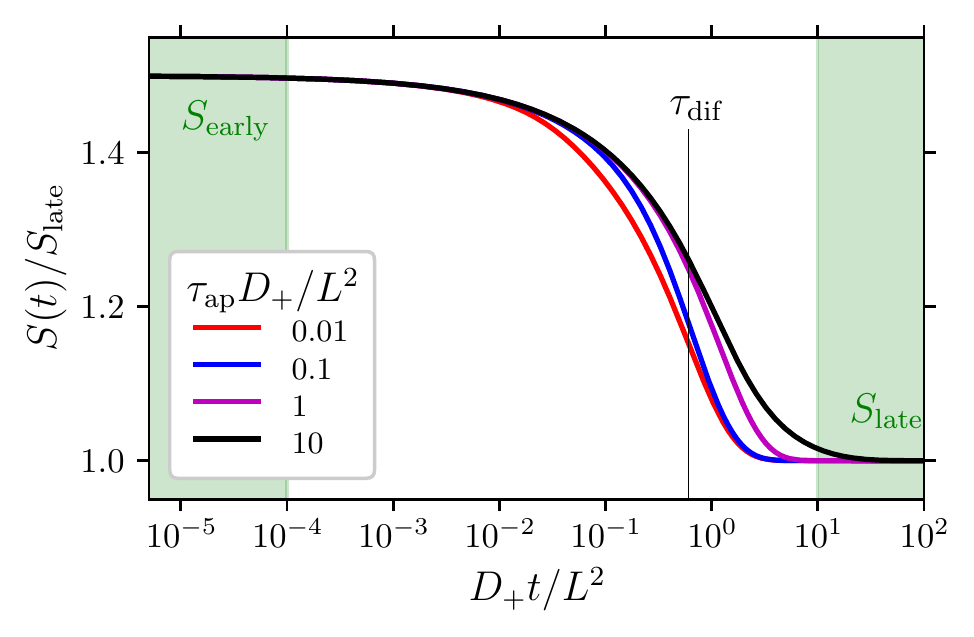}
\caption{$S(t)/S_\mathrm{late}$ [\cref{eq:Sttheory}] for 
for several $\tau_\mathrm{ap}D_{+}/L^2$ [$0.01$ (red), $0.1$ (blue), $1$ (magenta), $10$ (black)] with $\xi\equiv D_{+}/D_{-}=2$, $\alpha_{+}=0.5$, $\alpha_{-}=0.1$, $z_{+}=-z_{-}=1$ and $\max(j)=500$ throughout. For these parameters, $S_\mathrm{early}=(3/2)S_\mathrm{late}$.}
\label{fig:VTtau}
\end{figure}

We introduce two dimensionless parameters: the strength of the thermal gradient $\epsilon = \Delta T_{\infty}/T_{0}$ and the Debye separation parameter $n=L/\lambda $, with $\lambda=[e^2 (\rho_{+,0}z_{+}^{2}+\rho_{-,0}z_{-}^{2})/\varepsilon_{0}\varepsilon\kb T_{0}]^{1/2}$ being the usual Debye length. 
For $\epsilon \ll1$ and $n\gg1$ (in our experiments $n>3.5\times10^6$), \cref{eq:1dflux,eq:electrokinetic,eq:initandbc} can be solved analytically.
In \cref{sec:Sttheoryderivation} of the supplementary material we find the Seebeck coefficient 
\begin{align}\label{eq:Sttheory}
S(t) &=\frac{2(S_\mathrm{early}-S_\mathrm{late})}{1-\exp\left(-t/\tau_\mathrm{ap}\right)}
\sum_{j=1}^{\infty}\frac{{\mathrm{e}}^{- \mathcal{N}_{j}^2 t /(\mathcal{N}_{0}^2\tau_\mathrm{dif}) }-{\mathrm{e}}^{-t/\tau_\mathrm{ap} }}{ \mathcal{N}_{j}^{2} [1- \mathcal{N}_{j}^2 \tau_\mathrm{ap} /(\mathcal{N}_{0}^2\tau_\mathrm{dif})]}\nn
&\quad +S_\mathrm{late}+\mathcal{O}\left(n^{-1}, \epsilon\right)\,,
\end{align}
with $\mathcal{N}_{j}=(j-1/2)\pi$. 
Refining its above definition, $\tau_\mathrm{dif}= 4L^2/(\pi^2 D_{a})$ now contains the ambipolar salt diffusivity \cite{newman2012electrochemical},
\begin{equation}\label{eq:amphotericdiffusivity}
D_{a}=\frac{(z_{+}-z_{-})D_{+}D_{-}}{z_{+}D_{+}-z_{-}D_{-}}\,.
\end{equation}
We plot \cref{eq:Sttheory} in \cref{fig:VTtau} for several $\tau_\mathrm{ap}$ and see that $S(t)=S_\mathrm{early}$ for $t\ll\tau_\mathrm{ap}, \tau_\mathrm{dif}$ and $S(t)=S_\mathrm{late}$ for $t\gg\tau_\mathrm{ap}, \tau_\mathrm{dif}$.

\begin{table*}[t]
 \scriptsize
 \centering
 \setlength{\tabcolsep}{8pt}
 \begin{tabular}{ccccccccc}
\hline\hline \vspace{-2.5mm}\\ \vspace{0.5mm}
 electrolyte & {$z_{+}$} &{$z_{-}$} & $\rho_{s}\!$ (mM) & ${10^9} D_{+}~{(\si{\meter\squared\per\second})}$ & ${10^9} D_{-}~{(\si{\meter\squared\per\second})}$ & ${10^9} D_{a}~{(\si{\meter\squared\per\second})}$	 & ${10^{-3}} \tau_\mathrm{dif}^\mathrm{th}~(\si{\second})$ & ${10^{-3}} \tau_\mathrm{dif}^\mathrm{exp}~(\si{\second})$ \\\hline
 TBAOH	 		&{$1$}&{$-1$}	&5	&0.518	&5.280 	&0.943	&2.7   	&1.2		\\
 KCl	 			&{$1$}&{$-1$}	&5	&1.957	&2.033 	&1.994	&1.3 		&1.0		\\
 HCl	 			&{$1$}&{$-1$}	&2	&9.315 	&2.033 	&3.338	&0.76	&0.91	 	\\
 Li$_{2}$SO$_4$	&{$1$}&{$-2$}	&10	&1.028	&1.065	&1.040	&2.4		&1.5	 	\\
 MgSO$_4$		&{$2$}&{$-2$}	&10	&0.706	&1.065	&0.859	&3.0		&2.1 	 	\\
 \hline\hline
 \end{tabular}
 \caption{Properties of the aqueous electrolytes used here. All diffusivities at infinite dilution $D_{\pm}$ are taken from \cit{mills2013self}. 
 We found $D_{a}$ with \cref{eq:amphotericdiffusivity}. Next, for $\tau_\mathrm{dif}^\mathrm{th}=4L^2/(\pi^2D_{a})$ we used the electrode separation $2L=\SI{5.0}{\milli\meter}$ of our setup. 
A fitting procedure described in \cref{sec:results} yields the timescale $\tau_\mathrm{dif}^\mathrm{exp}$ with which the experimentally-measured thermovoltage $V_{T}(t)$ relaxed to its steady state.}
 \label{table1}
\end{table*}

\section{Experimental section}\label{sec:experiments}
\subsection{Setup and Procedure}\label{sec:setup}
We performed experiments with a homebuilt thermoelectric cell, sketched in \cref{fig:expsetup} and shown in more detail in \cref{figS3} of the supplementary material.
The cell comprised different electrolytes between two parallel, coaxial, disk-shaped titanium electrodes. 
The electrodes were \SI{5.0}{\milli\meter} thick and \SI{43.0}{\milli\meter} in diameter.
We stuck two Peltier elements (TES1 12704, dimensions: \SI{3 x 30 x 30}{\milli\metre}) to the outer sides of the electrodes and controlled them with a Neocera LTC21 PID temperature controller.
Type-K thermocouples and a digital multimeter (Minipa U1252A) measured the temperatures $T_{0}$ and $T_{1}$ of the bottom and top electrode.
The tips of these thermocouples were in contact with the electrodes through \SI{3}{\milli\meter}-deep, \SI{1.5}{\milli\meter}-wide holes drilled \SI{3}{\milli\meter} away from the edge of the Peltier elements.
A second multimeter (Minipa 8156A, impedance \SI{2.5e-9}{\ohm}) measured the voltage difference $\Delta V$ between the electrodes; a computer recorded $\Delta V$ automatically on three-second intervals.

The electrodes were held in a solid Teflon support, with a cylindrical cavity coaxial with the electrodes. 
The cavity was \SI{5.0}{\milli\meter} high and \SI{23.5}{\milli\meter} in diameter. 
Two O-rings between the electrodes and the Teflon block sealed off this cavity.
Two needles brought different electrolytes into the cavity through \SI{1}{\milli\meter} holes drilled laterally on opposite sides of the Teflon block.
We investigated aqueous solutions of TBAOH (tetrabutylammonium hydroxide), KCl (potassium chloride), HCl (hydrochloric acid), Li$_2$SO$_4$ (lithium sulfate), and MgSO$_4$ (magnesium sulfate).
We obtained all chemicals from Sigma-Aldrich and used them without further purification.
\Cref{table1} lists their ionic valencies $z_{\pm}$, salt concentrations $\rhos$, literature diffusivities $D_{\pm}$ at infinite dilution, ambipolar diffusivity $D_{a}$ [\cref{eq:amphotericdiffusivity}], predicted relaxation times $\tau_\mathrm{dif}^\mathrm{th}$, and experimental relaxation times $\tau_\mathrm{dif}^\mathrm{exp}$ as determined in \cref{sec:results}.
At small $\rhos$ ($\approx10$~mM), $D_{\pm}$ are often roughly $10\%$ smaller than at infinite dilution and $\tau_\mathrm{dif}^\mathrm{exp}$ $10\%$ larger, accordingly \cite{mills2013self}.
Moreover, for the $\rhos$ used here, Debye lengths are between $\lambda=\SI{1.5}{\nano\meter}$ for 10 mM $2:2$ and $\lambda=\SI{6.8}{\nano\meter}$ for 2 mM $1:1$ electrolytes, such that $n=L/\lambda>3.5\times10^6$.
We note that our setup is very similar to the one of \cit{Bonetti2015}, with the differences lying in the platinum foil electrodes and EMIMTFSI in acetonitrile electrolyte used there.

After filling the cavity with electrolyte, we measured a spontaneous potential difference (SPD) between the electrodes, even in the absence of imposed thermal gradients ($T_{0}=T_{1}=\SI{294}{\kelvin}$). 
This SPD probably stems from oxidation or reduction of the metallic electrode surfaces exposed to the liquid.
For the first round of experiments, we waited \SI{24}{\hour} for the SPD to stabilize within $\approx\SI{5}{\milli\volt}$ in \SI{1000}{\second}. 
For later experiments with other electrolytes, we waited $\approx\SI{2}{\hour}$ until the SPD stabilized to the same degree. 
For all electrolytes, the absolute value of the SPD was lower than \SI{100}{\milli\volt}.
After these waiting times, we started heating-cooling cycles: 
First, the top electrode warmed up to $T_{1}(t)=T_{0}+\Delta T(t)$ during a heating phase of \SI{e4}{\second}---three times longer than the largest $\tau_\mathrm{dif}^\mathrm{th}$. 
We used $\Delta T_{\infty}=\SI{11.63}{\kelvin}$ throughout.
Second, we brought $T_{1}$ back to $T_{0}=\SI{294}{\kelvin}$ during the cooling phase, also of \SI{e4}{\second}.
We repeated this procedure at least three times for each electrolyte. 
From our $\Delta V(t)$ measurements, we determined the thermovoltage by subtracting the SPD, $V_{T}(t)=\Delta V(t)-\Delta V(t_{0})$, where $t_{0}$ is the time that we start the heating-cooling cycles.

\subsection{Experimental Results}\label{sec:results}
\begin{figure}
\includegraphics[width=8.6cm]{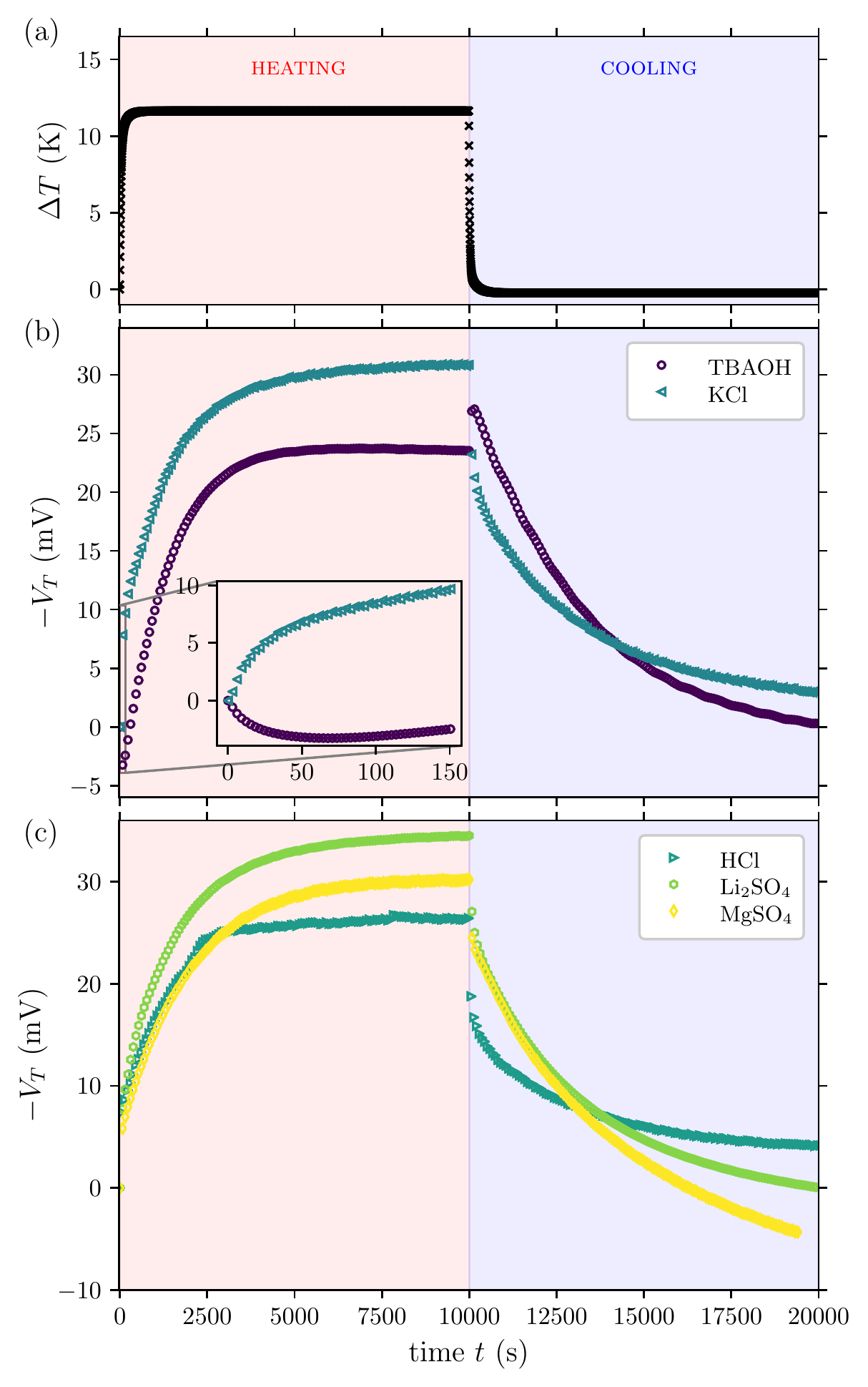}
\caption{The temperature difference $\Delta T(t)$ (a) and thermovoltage $V_{T}(t)$ [(b) and (c)] during one heating-cooling cycle.}
\label{fig:TandVT}
\end{figure}

\Cref{fig:TandVT} shows $\Delta T(t)$ (a) during a heating-cooling cycle and $V_{T}(t)$ [(b) and (c)] generated for the five different electrolytes during such a cycle.
First, we see that ${V_T}{(t)}$ changes rapidly for all electrolytes during the first \SI{2e3}{\second} of heating and cooling and slower during the \SI{8e3}{\second} thereafter. 
From the inset of \cref{fig:TandVT}(b), portraying the $0<t<\SI{150}{\second}$ range, we see that $V_{T}(t)$ varies non-monotonously for TBAOH. 
All electrolytes roughly reach a steady state within \SI{e4}{\second}. 

While \cref{fig:TandVT} presents data for one heating-cooling cycle for each electrolyte, to determine the characteristic timescale $\tau_\mathrm{dif}^\mathrm{exp}$ with which $V_{T}(t)$ increased, 
we averaged $V_{T}(t)$ over several (between three and five) heating-cooling cycles.
In calculating these averages, we reset $V_{T}(t)=0$ at the start of each cycle, to offset the \si{\milli\volt} irreversibilities seen in \cref{fig:TandVT} where $V_{T}\neq0$ for most electrolytes after a full cycle.
Likewise, \cref{figS4} of the supplementary material shows that $V_{T}$ typically differs a few mV during different cycles of the same electrolyte. 
We then numerically fit $V_{T}=V_{T,1}-V_{T,2}\exp{(-t/\tau_\mathrm{dif}^\mathrm{exp})}$ to the $t>\SI{100}{\second}$ data of the heating stage \cite{Note5}; \cref{table1} shows the fit parameters $\tau_\mathrm{dif}^\mathrm{exp}$. 
Discrepancies between $\tau_\mathrm{dif}^\mathrm{exp}$ and $\tau_\mathrm{dif}^\mathrm{th}$ are largest (a factor 2) for TBAOH.
The theory and experiments agree on HCl and KCl having the smallest $\tau_\mathrm{dif}$, and MgSO$_4$ and Li$_2$SO$_4$ having among the largest $\tau_\mathrm{dif}$. 

\begin{figure}
\includegraphics[width=8.6cm]{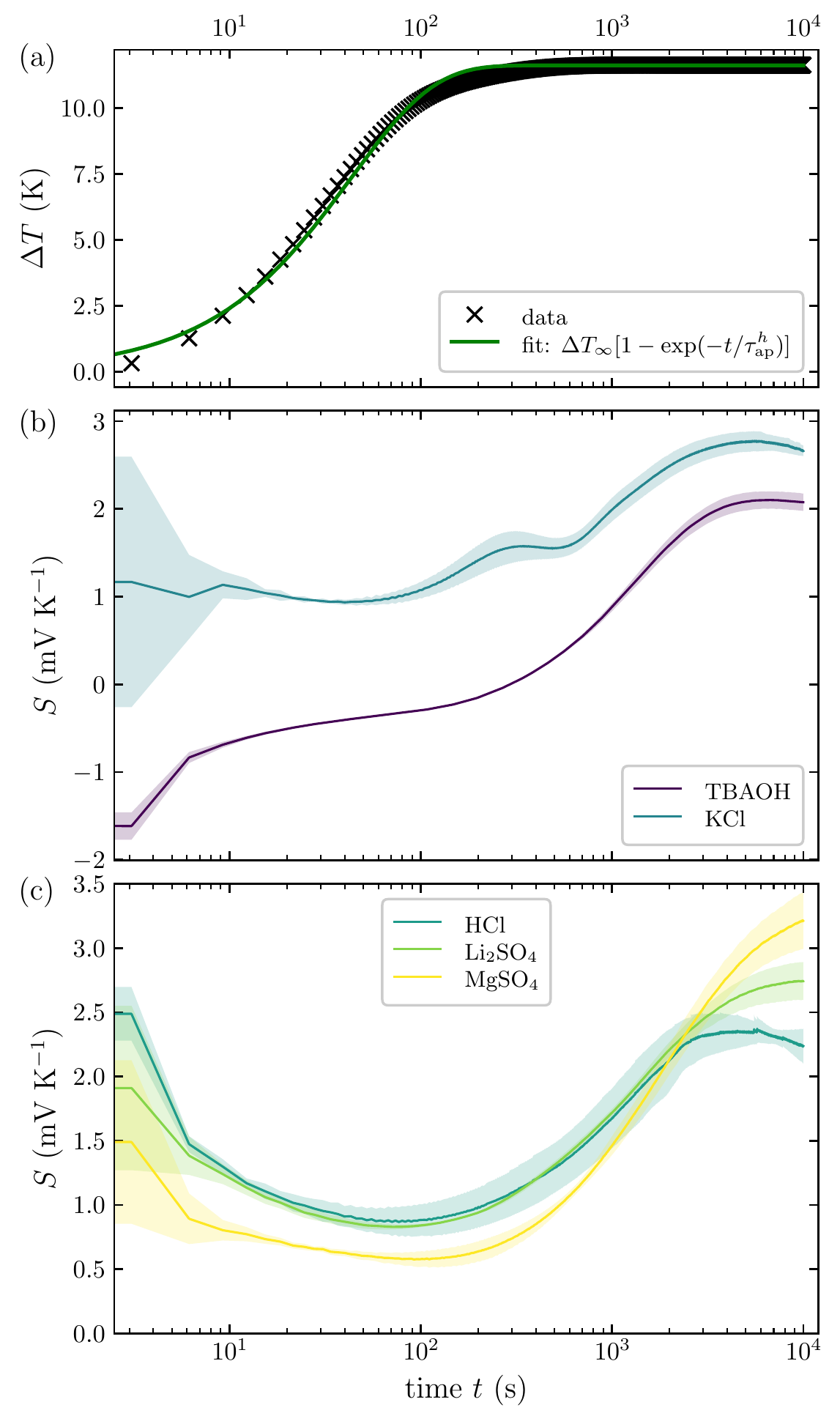}
\caption{Average Seebeck coefficient $S(t)$ (symbols) for several electrolytes [(b) and (c)] in response to a time-dependent temperature difference (a). 
Next to $\Delta T(t)$ (crosses), we also show the fitted function $\Delta T(t)=\Delta T_{\infty}\left[1-\exp (-t/\tau^{h}_\mathrm{ap})\right]$ (green line), where $\tau^{h}_\mathrm{ap}=\SI{43}{\second}$. 
The lines in (b) and (c) show $S(t)$ averaged over several runs and one standard deviation above and below the average with a shaded area.}
\label{fig:Sexp2}
\end{figure}
\Cref{fig:Sexp2}(a) shows the same heating-stage $\Delta T(t)$-data as \cref{fig:TandVT}(a) on semi-log scale.
This data is decently approximated by $\Delta T(t)/\Delta T_{\infty}=1-\exp (-t/\tau^{h}_\mathrm{ap})$ ({green} line) with the fit parameter $\tau^{h}_\mathrm{ap}=\SI{43}{\second}$.
During the cooling stage, $\Delta T(t)$ lowered slower: we found that $\Delta T(t)=\Delta T(0)\exp(-t/\tau^{c}_\mathrm{ap})$ with $\tau^{c}_\mathrm{ap}=\SI{59}{\second}$ then decently fits to the $\Delta T$ data (not shown). 
Note that both $\tau^{h}_\mathrm{ap}$ and $\tau^{c}_\mathrm{ap}$ are, indeed, larger than the timescale $\tau_{T}= \SI{18}{\second}$ with which the electrolyte should relax had we exposed it to a (theoretical) stepwise temperature difference. 
Next, we plot the time-dependent Seebeck coefficient $S(t)=-V_{T}(t)/\Delta T(t)$ (lines) in Figs.~\ref{fig:Sexp2}(b) and ~\ref{fig:Sexp2}(c). 
For each electrolyte, we calculated $S(t)$ using the above-described $V_{T}(t)$ averages and an interpolation through the $\Delta T(t)$-data of \cref{fig:Sexp2}(a).
Note that we can only plot $S(t)$ for the heating phase of the cycle.
 During the cooling phase, $\Delta T(t)$ approaches zero at late times. 
At late times, however, the small irreversibilities yielding $V_{T}\neq0$ would cause $S(t)$ to diverge, making $S(t)$ an impractical measure of the small measured potential difference.
To show the spread in $S(t)$ between different cycles, the shaded areas in Figs.~\ref{fig:Sexp2}(b) and ~\ref{fig:Sexp2}(c) indicate one standard deviation below and above the average $S(t)$.
We see there that data differ most for $t<\SI{10}{\second}$. 
This is probably because, at these early times, $S(t)$ results from the division of two numbers ($V_{T}$ and $\Delta T$) whose absolute values are small compared to the error in their measurement. 
At later times, until about $t=\SI{100}{\second}$, $S(t)$ is stable, then goes through a transition period, and finally reaches a second plateau around $t=\SI{5000}{\second}$.

\section{Discussion}\label{sec:Discussion}
\subsection{Seebeck coefficient literature comparison}
With literature $Q_{i}^{*}$ data from \cit{agar1989}, we determined predictions for $S_\mathrm{late}$ and $S_\mathrm{early}$ [\cref{eq:Seebecklate,eq:Seebeckearly}] and plotted them in \cref{figS5} of the supplementary material.
That figure suggests that $\mathrm{sgn}(S_\mathrm{early})=\mathrm{sgn}(S_\mathrm{late})$ for most electrolytes, TBAOH being a notable exception.
Indeed, we observe $S(t)$-sign switching for TBAOH in our experimental data of \cref{fig:Sexp2}. 
Other than this qualitative agreement, there is a clear quantitative difference between our \cref{fig:Sexp2} and the literature predictions in \cref{figS5} of the supplementary material{:}
Our measured $S_{\rm late}\approx \SI{2}{\milli\volt\per\kelvin}$ for many electrolytes is larger than expected on the basis of \cit{agar1989}.

Let us speculate about the possible oversimplifications in our theory that could have caused this discrepancy.
First, {our model accounts only for $\mathcal{O}(\epsilon)$ effects and is thus, strictly speaking, only valid for vanishingly small $\epsilon=\Delta T_{\infty}/T_{0}$ (the $\epsilon\approx 0.03$ of our experiments is similar to other electrolyte-Seebeck literature \cite{Bonetti2011, Bonetti2015}).
At $\mathcal{O}(\epsilon^2)$, neglected by us, one should account for $c(x,t)$ and $T(x,t)$-dependence of the parameters $D_{i}$, $Q^{*}_{i}$, and $\varepsilon$.}
Moreover, the measured $V_{T}\approx\SI{20}{\milli\volt}$ suggest that the Debye-Falkenhagen-type equations of the supplementary material [cf.~\cref{eq:3}]{---to which \cref{eq:1dflux,eq:electrokinetic,eq:initandbc} simplify when the local electrostatic potential is small} $\psi\ll e/(\kb T_{0})\approx \SI{24}{\milli\volt}$ \cite{janssen2018}---are stretched beyond their region of validity. 
For larger $\epsilon$, leading to larger local electrostatic potentials, the finite size of ions becomes increasingly important. 
Hence, future work could extend our model with a Stern layer or with a modified Nernst-Planck expression for the ion fluxes \cite{kilic2007}.
Second, we assumed the thermal gradient to have no lateral component. 
In our experiments, the electrolyte temperature could have varied slightly laterally near the edge of the cavity, as the Teflon and O-rings, parts of low thermal conductivity, {were} in thermal contact with room-temperature air.
Third, we did not account for fluid flow in our model.
Had we heated the bottom plate rather than the top plate by the same $\Delta T_{\infty}$, this simplification would not have been allowed.
With parameters typical for water, our setup would have been characterised by a Rayleigh number $\textrm{Ra}=\SI{2.1e4}{}$.
This is an order of magnitude larger than the critical value $\textrm{Ra}_c=1708$ beyond which a Rayleigh-Benard instability would have stirred the electrolyte \cite{leal2007advanced}.
However, in our setup with a hotter top electrode, ignoring convection seems warranted \cite{Salazar2014, Bonetti2015}.
Fourth, our mesoscopic model lumps all the solvent's (water) electrostatic properties into a single parameter (the dielectric constant). 
Hence, our model does not account for specific ion-solvation effects \cite{levin2009ions}, nor does it account for thermoelectric fields generated by water polarization in as much detail as \cit{dilecce2018}. 
Fifth, we ignored electrostatic edge effects, electronic surface charge, and image charge effects.
In reality, the cavity aspect ratio was $5.0/23.5=0.21$ and nA currents could have flowed through the external circuit and \SI{2.5e9}{\ohm} multimeter to measure the thermovoltage ($\sim\SI{20}{\milli\volt}$).
To account for image charge effects, one should generalize the methods of, for instance, \cit{kjellander1985} to out-of-equilibrium and nonisothermal systems.
Finally, a better understanding of specific electrolyte-electrode interactions, including ionic adsorption \cite{bickel2014micro} and redox reactions \cite{deandrade2016,Note6} is desirable.
We observed no changes in roughness, color, or reflection on the surfaces of our Ti electrodes, even after months of usage. 
Moreover, Ti appears high in the Galvanic series of metals in contact with flowing seawater \cite{rodriguez2014anticipated}.
These observations and data support our simplifying assumption that no Faradaic reactions took place between the electrodes and the electrolyte.
Theoretical studies have shown that different ion adsorption and desorption rates on electrodes can cause a potential difference of around the thermal voltage \cite{Bousiadi2018}.
Hence, specific ion adsorption may have contributed to our measured SPD. 

It would be interesting to extend our theoretical model to account for all the above phenomena---or to perform further experiments with setups that better respect our current assumptions.
Additional parameters such as the adsorption affinity and reaction entropy of redox couples---different for each temperature, salinity, and electrolyte-electrode combination---would need to be measured, with impedance spectroscopy \cite{okazaki2019} and cyclic voltammetry, for instance.

\subsection{Determining single-ion Soret coefficients from the time-dependent Seebeck coefficient}
The above-mentioned literature {values for} $Q_{i}^{*}$ essentially all derive from Soret effect experiments that measured the \textit{sum} $\alpha_{+}+\alpha_{-}$ \cite{agar1957thermal, Breck1957, Agar1963, blokhra1972, lin1973heats, Petit1987}.
From these data, the single-ion Soret coefficients $\alpha_{+}$ and $\alpha_{-}$ were determined either by (arbitrarily) setting $\alpha_\mathrm{Cl}=0$ as a reference point \cite{Agar1963} or by the ``reduction rule" of Takeyema and Nakashima \cite{Takeyama1988}.
To the best of our knowledge, the \textit{individual} $\alpha_{\pm}$ were never directly determined---not by potentiometric experiments nor by any other method.
Yet, our theoretical expression \cref{eq:Sttheory} for $S(t)$ gives us the opportunity to do precisely this.
	
\Cref{eq:Sttheory} depends parametrically on $\alpha_{\pm}, D_{\pm}$, and $\tau_\textrm{ap}$.
Hence, with enough experimental data, we should in principle be able to determine all five parameters through fitting \cref{eq:Sttheory} to data.
Instead of doing so, we choose to use our independent measurement of $\tau_\textrm{ap}=\SI{43}{\second}$ and literature values for $D_{\pm}$ (\cref{table1}) \cite{Note7}. 
We then determined the two single-ion Soret coefficients $\alpha_{\pm}$ of all five binary electrolytes from a two-parameter fit of \cref{eq:Sttheory} to the average $S(t)$ data shown in \cref{fig:Sexp2}(b). 
As we wanted to fit the complete transient behavior of $S(t)$, for each electrolyte we picked 60 approximately-logarithmically-separated data points. 
In this way, we prevented over-representing the late-time behavior of $S(t)$, for which we have much more data. 
With the {\sc curve$\_$fit} algorithm of {\sc scipy}, we {found} the $\alpha_{\pm}$ values of the second column of \cref{table2}. 
\begin{table}
\scriptsize
\centering
\setlength{\tabcolsep}{8pt}
	\begin{tabular}{lcccc}
		\hline\hline\vspace{0.1cm}
		&\hspace{1cm} $\alpha_{i}$\\
		ion		& \cref{eq:Sttheory}			& \cref{eq:alphapm}	&\cit{Takeyama1988}	\\ \hline \vspace{0.05cm}
		H$^{+}$	 						&	-0.4			 	&0.8							&2.72							\\
		Li$^{+}$ 							&	1069.1			&919.7						&0.11							\\
		K$^{+}$							& 	611.4			&569.6						&0.53 							\\
		Mg$^{2+}$						& 	203.7			&182.9						&1.89							\\
		TBA$^{+}$ 						&	37.6				&30.2						&3.95							\\
		OH$^{-}$ 							&	7.7				&5.8							&3.52 							\\
		Cl$^{-}$(K$^{+}$)					&	579.2			&537.4						&0.11	 						\\
		Cl$^{-}$(H$^{+}$) 					&	-25.5				&-28.1						&0.11 							\\
		SO$_{4}^{2-}$(Li$^{+}$)\vspace{0.5mm} 	&	1019.2			&872.0						& 								\\
		SO$_{4}^{2-}$(Mg$^{2+}$)\vspace{0.5mm}&	127.6			&108.7						&								\\
		\hline\hline
	\end{tabular}
	\caption{Single-ion Soret coefficients determined from our experimental $S(t)$ data with \cref{eq:Sttheory,eq:alphapm}.
	Chloride and sulfate ions are in two different electrolytes that we considered. Accordingly, between brackets we mention the other ion present in solution. 
	The right column was determined with $\alpha_{i}=Q_{i}^{*}/(2\kb T)$ using $T=\SI{294}{\kelvin}$ and ionic heat of transport $Q_{i}^{*}$ data from \cit{Takeyama1988}.
	When \cit{Takeyama1988} mentions multiple sources, we pick the first entry of the infinite dilution column. }\label{table2}
\end{table}

For comparison, we also determined $\alpha_{\pm}$ from $S(t)$ in a simpler way, as follows. 
We note that \cref{eq:Seebecklate,eq:Seebeckearly} can be rewritten to two equations for $\alpha_{+}$ and $\alpha_{-}$,
\begin{equation}\label{eq:alphapm}
\frac{e\alpha_{\pm}}{2\kb}= \frac{D_{+} z_{+}-D_{-}z_{-}}{D_{+}-D_{-}}S_\mathrm{early}-\frac{D_{\mp}(z_{+}-z_{-})}{D_{+}-D_{-}}S_\mathrm{late}\,.
\end{equation}
In \cref{fig:VTtau} we saw that $S(t\ll \tau_\mathrm{dif})\approx S_\mathrm{early}$ and $S(t\gg \tau_\mathrm{dif})\approx S_\mathrm{late}$. 
Accordingly, for all electrolytes, we determined $S_\mathrm{early}$ as the average of the first 30 of our \mbox{3-s-separated} data points and $S_\mathrm{late}$ from the maximum of $S(t)$. 
Inserting the thus-found $S_\mathrm{late}$ and $S_\mathrm{early}$ and $D_{\pm}$ from \cref{table1} into \cref{eq:alphapm} yields $\alpha_{\pm}$ as listed in the third column of \cref{table2}.
We see that \cref{eq:Sttheory,eq:alphapm} yield comparable $\alpha_{\pm}$ for the same experiments, even though they use different subsets of the same $S(t)$ data.
Neither of these equations, however, give consistent $\alpha_{\pm}$ values for the same ion measured in different electrolytes. 
While, indeed, $\alpha_{\pm}$ may differ with $T$, $\rhos$, and electrolyte composition \cite{dilecce2017}, the spread in $\alpha_{-}$ of Cl$^{-}$ {and SO$_{4}^{2-}$} is unrealistically large.

Finally, \cref{table2} contains $\alpha_{i}=Q_{i}^{*}/(2\kb T)$ calculated with $Q_{i}^{*}$ at infinite dilution from \cit{Takeyama1988}. 
Previous works accounted for the $\rhos$-dependence of $Q_{i}^{*}$ with a correction factor close to 0.95 \cite{Agar1960, Leaist1994}. 
Clearly, this subtlety cannot have caused the large discrepancies that we observe in \cref{table2} between our and literature data.
While the individual $\alpha_{\pm}$ have not been measure before, the sum $\alpha_{+}+\alpha_{-}$ \textit{has} been accurately determined.
For all electrolytes considered, $\alpha_{+}+\alpha_{-}$ should be of order 1, which is a consistency check not satisfied by our $\alpha_{\pm}$ values.

Obviously, the above discrepancies must in part be due to the larger-than-expected $S(t)$ discussed above.
Apart from that, it seems that \cref{eq:Sttheory,eq:alphapm} are inherently less accurate for electrolytes with similar cationic and anionic diffusivities:
$\alpha_{\pm}$ is huge for the constituent ions of KCl and Li$_{2}$SO$_{4}$ (and MgSO$_{4}$, to a lesser extent), for which we find $D_{+}/D_{-}=0.96$ and $0.97$ with the data for $D_{\pm}$ of \cref{table1}.
On the basis of \cref{eq:Seebecklate,eq:Seebeckearly}, one expects $S_\mathrm{late}=S_\mathrm{early}$ for such electrolytes.
While \cref{fig:Sexp2} shows that the difference between $S_\mathrm{late}$ and $S_\mathrm{early}$ for those electrolytes is, indeed, smaller than for all other electrolytes shown, it is also clear that $S_\mathrm{late}\neq S_\mathrm{early}$.
Even for experimental systems perfectly described by our theoretical model, \cref{eq:alphapm} for $\alpha_{\pm}$ is probably not accurate when $D_{+}\approx D_{-}$, as the term $1/(D_{+}-D_{-})$ becomes very sensitive to uncertainty (and $\rhos$-dependence) in the ionic diffusivities.
Interestingly, the aforementioned works \cite{agar1957thermal, Breck1957, blokhra1972, lin1973heats, Petit1987} that determined $\alpha_{+}+\alpha_{-}$ from $S_\mathrm{early}-S_\mathrm{late}$ did not suffer from the same diverging factor, probably because these authors included into \cref{eq:Seebecklate} only the cationic species with which their electrodes reacted electrochemically.
De Groot's \cite{de1947thermodynamique} Eqs.~(74) and (78) and Haase's \cite{haasse1953} Eqs. (17) and (22), however, are equivalent to our \cref{eq:Seebecklate,eq:Seebeckearly}. 
Thus, had they expressed $Q_{i}^{*}$ in terms of $V(t)$, their expressions would have suffered from the same divergence at $D_{+}=D_{-}$. 
 
\section{Conclusions}
We studied the transient thermovoltage $V_{T}(t)$ and Seebeck coefficient $S(t)=-V_{T}(t)/\Delta T(t)$ generated by a binary multivalent electrolyte subject to a time-dependent temperature difference $\Delta T(t)$. 
In particular, we rigorously rederived a theoretical literature expression for the ``instantaneous" Seebeck coefficients $S_\mathrm{early}$ [\cref{eq:Seebeckearly}] without enforcing local charge neutrality. 
Next, we performed experiments with five binary electrolytes near Ti electrodes.
To the best of our knowledge, no prior thermovoltage measurements were performed with such electrodes before.
We found steady-state Seebeck coefficients around $S_\mathrm{late}\approx \SI{2}{\milli\volt\per\kelvin}$ for several electrolytes, which is roughly one order of magnitude \textit{larger} than what was to be expected from inserting literature $Q_{i}^{*}$ data into \cref{eq:Seebecklate}.
While it is known that the steady-state coefficient $S_\mathrm{late}$ can change its sign upon changing temperature or salinity \cite{Roemer2013,dilecce2017}, in this work we showed that $S(t)$ can also switch sign \textit{in time}.
Taking prior diffusivity $D_{i}$ and heat of transport $Q_{i}^{*}$ data as input, our theory suggested that TBAOH would show such a sign reversal, which we, indeed, observed experimentally. 
We also found qualitative agreement between our theory and experiments for the timescale $\tau_\mathrm{dif}$ at which $S(t)$ transitions from $S_\mathrm{early}$ to $S_\mathrm{late}$.

Much recent literature on the electrolyte Seebeck effect aims at thermal energy harvesting applications, seeking electrode-fluid combinations that maximize $S(t)$, for instance with textured electrodes \cite{Salazar2014a, Bonetti2015, Zhao2016, Zhang2017}, nonaqueous electrolytes \cite{Bonetti2011,Zinovyeva2014, kim2018, han2020}, and nanoparticles \cite{putnam2005, Vigolo2010, huang2015, Sehnem2015, Salez2017, bhattacharya2020structural}. 
To put such studies in context requires detailed knowledge of dilute electrolytes between flat electrodes of various materials.
Yet, the accepted literature values of $Q_{\pm}^{*}$ (and equivalently $\alpha_{\pm}$) almost all trace back---via Takeyema and Nakashima's ``reduction rule" \cite{Takeyama1988}---to the Soret effect measurements of Agar and coworkers of the 1960s.
Indeed, all literature $\alpha_{\pm}$ values were determined with experiments that measured the \textit{sum} $\alpha_{+}+\alpha_{-}$.
We proposed and tested a new way of determining \textit{individual} $\alpha_{\pm}$ values by a two-parameter fit of \cref{eq:Sttheory} to experimental $S(t)$ data.
Our measured $S(t)$ already being larger than expected, the $\alpha_{\pm}$ that we found with this fitting procedure were also generally larger than literature values.
As our $V_{T}(t)$ measurements may have been caused in part by specific electrode-electrolyte interactions not covered in our current model, our derived $\alpha_{\pm}$ values are less reliable than those obtained by traditional methods.
To refine our method would require to relax (some of) the assumptions in our theoretical model, or to build experimental setups that respect these assumptions better. 
Thereafter, it would be interesting to generalize our method of determining $\alpha_{\pm}$ to more complex systems.
Introducing a third charged species---nanoparticles \cite{putnam2005,Salez2017} or colloids \cite{Vigolo2010,majee2011,Reichl2014a,huang2015}, for instance---will introduce an additional diffusivity and lengthscale(s), which may transpire into a three- rather than two-step $S(t)$-relaxation. 
Unfortunately, generalizing our derivation of the supplementary material to an $n$-component mixture seems prohibitively tedious.
Until then, it is tempting to assume that the derivations of multicomponent generalizations of \cref{eq:Seebecklate,eq:Seebeckearly} \cite{de1947thermodynamique,Agar1963,Bonetti2011,Zinovyeva2014,huang2015,Salez2017,salez2018,bhattacharya2020structural}, relying on unjustified local charge neutrality assumptions, led to correct results in that case as well.

\section*{supplementary material}
See the supplementary material for a derivation of \cref{eq:Seebecklate,eq:Sttheory}; for a detailed schematic of our experimental setup; for a figure of $V_{T}(t)$ for several heating-cooling cycles of TBAOH, KCl, and Li$_{2}$SO$_4$; and for a figure showing $S_\mathrm{late}$ vs. $S_\mathrm{early}$ using $Q_{i}^{*}$ data of \cit{agar1989}.

\section*{Authors' contribution}
A.L.S. conducted the experiments. 
A.L.S. and M.J. discussed and analyzed the experimental results. 
M.J. derived the theory and wrote the article. 
A.L.S. contributed to the editing thereof.

\section*{Acknowledgements}
We thank J. C. Everts, S. Kondrat, {S. Poulain}, and A. W\"{u}rger for insightful comments on this manuscript and M. Bonetti, S. Nakamae, and M. Roger for inspiring discussions.
M.J. moreover acknowledges H. Stenmark and A. Carlson for support.
A.L.S. acknowledges financial support from research funding agencies CAPES (\textit{Coordenação de Aperfeiçoamento de Pessoal de N\'{i}vel Superior} - 88881.133118/2016-01), Cnpq (\textit{Conselho Nacional de Desenvolvimento Científico e Tecnológico} - 465259/2014-6), FAPESP (\textit{Fundação de Amparo à Pesquisa do Estado de São Paulo} - 2014/50983-3; 2016/24531-3), and INCTFCx (\textit{Instituto Nacional de Ciência e Tecnologia de Fluidos Complexos}).

\section*{Data AVAILABILITY STATEMENT}
The data that support the findings of this study are available from the corresponding author upon reasonable request.

\bibliographystyle{apsrev4-2}

\clearpage

\onecolumngrid
\begin{center}
{\bf SUPPLEMENTARY MATERIAL to: On the time-dependent electrolyte Seebeck effect}\\\vspace{0.3cm}
Andr\'{e} Luiz Sehnem and Mathijs Janssen
\end{center}
\twocolumngrid

\vspace{\columnsep}
\setcounter{equation}{0}
\setcounter{figure}{0}
\setcounter{section}{0}
\renewcommand{\theequation}{{S}\arabic{equation}}
\renewcommand{\thefigure}{{S}\arabic{figure}}
\definecolor{nr0}{rgb}{0.267004, 0.004874, 0.329415}
\definecolor{nr1}{rgb}{0.229739, 0.322361, 0.545706} 
\definecolor{nr2}{rgb}{0.127568, 0.566949, 0.550556} 
\definecolor{nr3}{rgb}{0.369214, 0.788888, 0.382914} 
\definecolor{nr4}{rgb}{0.993248, 0.906157, 0.143936} 

\section{Derivation of $S(t)$ [Eq.~(7)]}\label{sec:Sttheoryderivation}
In \cref{sec:dimless,sec:solinsdomain,sec:theorytdomain} we generalize the findings of \cit{Janssen2019} for a $1:1$ electrolyte subject to an instantaneous thermal gradient $T(x)=T_{0}+\Delta T (x+L)/(2L)$ to a $z_{+}:-z_{-}$ electrolyte.
In \cref{sec:slowDeltaT} we replace the $T(x)$ assumption with a time-dependent temperature difference of the form $T(x,t)=T_{0}+\Delta T_{\infty} \left[1-\exp (-t/\tau_\mathrm{ap})\right] (x+L)/(2L)$.

\subsection{Dimensionless formulation}\label{sec:dimless}
For the instantaneous thermal gradient $T(x)$, rather than \cref{eq:1dflux} of the main text, we have
\begin{equation}\label{eq:1dfluxinst}
-\frac{J_{i}}{D_{i}}=\partial_{x} \rho_{i}+\frac{z_{i}e\rho_{i}}{\kb T}\partial_{x} \psi +\frac{\rho_{i}Q^{*}_{i}}{\kb T^2}\frac{\Delta T}{2L}\,.
\end{equation}
Introducing the dimensionless observables $\tilde{t}=tD_{+}/L^2$, $\tilde{x}=x/L$, $\tilde{\rho}_{i}=\rho_{i}/\rhos $, $\tilde{q}=q/\rhos $, $\tilde{\psi}=e\psi/(\kb T_{0})$, $\tilde{T}=T/T_{0}$, and $\tilde{J}_{i}=J_{i}L/(D_{+}\rhos)$ we rewrite \cref{eq:1dfluxinst,eq:electrokinetic} of the main text to
\begin{subequations}\label{eq:dimensionlessequations}
\begin{align}
\partial^{2}_{\tilde{x}} \tilde{\psi}&=- n^{2}\frac{z_{+}\tilde{\rho}_{+}+z_{-}\tilde{\rho}_{-} }{z_{+}^{2}\tilde{\rho}_{+,0}+z_{-}^{2}\tilde{\rho}_{-,0}}\,,\label{eq:Poissondimless}\\
\partial_{\tilde{t}} \tilde{\rho}_{+}&=\partial_{\tilde{x}}\left(\partial_{\tilde{x}} \tilde{\rho}_{+}+\frac{z_{+}\tilde{\rho}_{+}}{\tilde{T}}\partial_{\tilde{x}} \tilde{\psi} \right)\,,\label{eq:NPdimless}\\
\xi \partial_{\tilde{t}} \tilde{\rho}_{-}&=\partial_{\tilde{x}}\left(\partial_{\tilde{x}} \tilde{\rho}_{-}+\frac{z_{-}\tilde{\rho}_{-}}{\tilde{T}}\partial_{\tilde{x}} \tilde{\psi} \right)\,,\label{eq:NPdimless2}
\end{align}
\end{subequations}
with $\xi=D_{+}/D_{-}$ the diffusivity ratio.
In \cref{eq:dimensionlessequations} we omitted terms proportional to $\partial_{\tilde{x}}D_{i}\partial_{\tilde{x}} \tilde{\rho}_{i}$ and to $\partial_{\tilde{x}}\tilde{\rho}_{i}\Delta T/T$ as they are of subleading importance in the small-temperature limit that we consider below.

In terms of the above dimensionless parameters, \cref{eq:initandbc} of the main text becomes
\begin{subequations}\label{eq:bcs}
\begin{align}
\tilde{\rho}_{i}(\tilde{x},\tilde{t}=0)&=\tilde{\rho}_{i,0}\,,\\
\tilde{q}_{0}(\tilde{x},\tilde{t}=0)&=\tilde{\rho}_{+,0}z_{+}+\tilde{\rho}_{-,0}z_{-}=0\,,\\
\partial_{\tilde{x}}\tilde{\psi}(\pm 1,\tilde{t}\,)&=0 \,, \\
\tilde{J}_{\pm}(\pm 1,\tilde{t}\,)&=\partial_{\tilde{x}} \tilde{\rho}_{\pm}+\frac{\alpha_{\pm}\tilde{\rho}_{\pm}\epsilon}{\tilde{T}}=0\,,\label{eq:nofluxdimless}
\end{align}
\end{subequations}
where $\epsilon=\Delta T/T_{0}$ sets the strength of the thermal gradient and where $\tilde{\rho}_{i,0}$ are stochiometric coefficients that count the number of cations and anions into which a single salt molecule dissociates in solution.
For Li$_{2}$SO$_{4}$, for example, $\tilde{\rho}_{+,0}=2, \tilde{\rho}_{-,0}=1, z_{+}=1$, and $z_{-}=-2$. 

From hereon, we focus on small applied temperature differences $\epsilon\ll1$ and expand all observables and parameters in $\epsilon$, using the general notation $f=f_{0}+\epsilon f_{1}+\mathcal{O}\left(\epsilon^{2}\right)$. 
The unperturbed density and temperature profiles $\rho_{i,0}$ and $T_{0}$ as used above are in line with this definition. 
Using that $\tilde{\psi}_{0}=0$ and $\tilde{T}_{0}=1$, we find that the terms of lowest order in $\epsilon$ of \cref{eq:dimensionlessequations} are at $\mathcal{O}\left(\epsilon\right)$.
They read 
\begin{subequations}
\begin{align}
\partial^{2}_{\tilde{x}} \tilde{\psi}_{1}&=- n_{0}^{2}\frac{z_{+}\tilde{\rho}_{+,1}+z_{-}\tilde{\rho}_{-,1} }{z_{+}^{2}\tilde{\rho}_{+,0}+z_{-}^{2}\tilde{\rho}_{-,0}}\,,\label{eq:Poissondimless_b}\\
\partial_{\tilde{t}} \tilde{\rho}_{+,1}&=\partial_{\tilde{x}}\left(\partial_{\tilde{x}} \tilde{\rho}_{+,1}+z_{+}\tilde{\rho}_{+,0}\partial_{\tilde{x}} \tilde{\psi}_{1} \right)\,,\label{eq:NPdimless_b}\\
\xi_{0} \partial_{\tilde{t}} \tilde{\rho}_{-,1}&=\partial_{\tilde{x}}\left(\partial_{\tilde{x}} \tilde{\rho}_{-,1}+z_{-}\tilde{\rho}_{-,0}\partial_{\tilde{x}} \tilde{\psi}_{1} \right)\,.\label{eq:NPdimless2_b}
\end{align}
\end{subequations}
Note that we ignored $T$-dependence of $D_{\pm}$ in \cref{eq:NPdimless,eq:NPdimless2}---giving terms $\partial_{\tilde{x}}D_{i}\partial_{\tilde{x}} \tilde{\rho}_{i,1}=\mathcal{O}\left(\epsilon^2\right)$ that would not have appeared in \cref{eq:NPdimless_b,eq:NPdimless2_b}. 
Moreover, in $n_{0}$ appears the first term of a small-$\epsilon$ expansion of the dielectric constant $\varepsilon$, though we do not use $\varepsilon_{0}$ here as this symbol is reserved for the vacuum permittivity.
Instead of $n_{0}$, $\xi_{0}$, and $\alpha_{i,0}$, for notational convenience, we write $n$, $\xi$, and $\alpha_{i}$ from hereon.
Practically, the above small-$\epsilon$ expansion has simplified our model to the point that we can ignore the $c(x,t)$ and $T(x,t)$-dependence of the parameters $D_{i}$, $Q^{*}_{i}$, and $\varepsilon$. Strictly speaking, however, this simplified model is only applicable for infinitesimally small $\Delta T/T_{0}$.

Inserting \cref{eq:Poissondimless_b} into \cref{eq:NPdimless_b,eq:NPdimless2_b}, using initial charge neutrality $z_{+}\tilde{\rho}_{+,0}+z_{-}\tilde{\rho}_{-,0}=0$, and writing $\chi=z_{+}/z_{-}$, we find
\begin{subequations}\label{eq:3}
\begin{align}
\partial_{\tilde{t}} \tilde{\rho}_{+,1}&=\partial^2_{\tilde{x}} \tilde{\rho}_{+,1}-n^2\frac{\chi \tilde{\rho}_{+,1}+ \tilde{\rho}_{-,1}}{\chi-1} \,,\\
\xi\partial_{\tilde{t}} \tilde{\rho}_{-,1}&=\partial^2_{\tilde{x}} \tilde{\rho}_{-,1}- n^2\frac{\chi\tilde{\rho}_{+,1}+ \tilde{\rho}_{-,1}}{1-\chi}\,.
\end{align}
\end{subequations}

\subsection{Solution for the thermovoltage in the $s$ domain}\label{sec:solinsdomain}
We apply Laplace transformations to both sides of \cref{eq:3} [for a function $f(x,t)$ we write $\hat{f}(x,s)=\int_{0}^{\infty}\dif \tilde{t}\,\exp{(-s\tilde{t}\,)}f(x,\tilde{t}\,)$] and group the result in a matrix equation,
\begin{equation}\label{eq:matrixequation}
 \begin{bmatrix}\begin{array}{c} \partial^2_{\tilde{x}}\hat{\tilde{\rho}}_{+,1}\\ \partial^2_{\tilde{x}}\hat{\tilde{\rho}}_{-,1} \end{array} \end{bmatrix} 
 = \def\arraystretch{2.1}\begin{bmatrix} 
	\displaystyle{s+ \frac{\chi n^2}{\chi-1}}&
	\displaystyle{  \frac{n^2}{\chi-1}} \\ 
	\displaystyle{  \frac{\chi n^2}{1-\chi}} &
	\displaystyle{\xi s+\frac{n^2}{1-\chi}} 
 \end{bmatrix} 
 \def\arraystretch{1}\begin{bmatrix} \begin{array}{c} \hat{\tilde{\rho}}_{+,1}\\ \hat{\tilde{\rho}}_{-,1} \end{array} \end{bmatrix}\,,
\end{equation}
which we write as $X''=MX$, with double primes indicating second partial derivatives on the vector $X=[\hat{\tilde{\rho}}_{+,1}, \hat{\tilde{\rho}}_{-,1}]^{\intercal}$.
$M$ is diagonalized by $M=PDP^{-1}$, where 
\begin{equation}
P=\begin{bmatrix} \nu_{1}&\nu_{2} \\ 
			1 &1 \end{bmatrix} \,,
			\qquad
D=\begin{bmatrix} \mu^2&0 \\ 
			0 &\eta^2 \end{bmatrix} 	\,,		
\end{equation}
and where $\nu_{1}, \nu_{2},\mu$, and $\eta$ read
\begin{subequations}\label{eq:nurandm}
\begin{align}
\nu_{1} &= \frac{ s (\xi-1)+ \zeta}{n^{2}}\frac{\chi -1}{2\chi }- \frac{\chi+1}{2\chi}\,, \\
\nu_{2} &= \frac{ s (\xi-1)-\zeta}{n^{2}} \frac{\chi-1 }{2\chi } - \frac{\chi+1}{2\chi}\,,\\
\mu^2 &=\frac{1}{2}\left[n^{2}+s(1+\xi)-\zeta\right] \,,\label{eq:mu}\\
\eta^2 &=\frac{1}{2}\left[n^{2}+s(1+\xi)+\zeta\right]\,,
\end{align}
\end{subequations}
respectively, with
\begin{equation}\label{eq:zeta}
\zeta=\sqrt{n^4+2 n^2 s(1-\xi) \frac{\chi+1}{\chi-1}+s^2(1-\xi)^2}\,.
\end{equation}
With $U=[u_{1}, u_{2}]^{\intercal}\equiv P^{-1}X$ we rewrite $X''=MX$ to \mbox{$U''=DU$}, which is solved by $u_{1}=a_{1}\sinh \mu \tilde{x}$ and $u_{2}=a_{2}\sinh \eta \tilde{x}$, with $a_{1}, a_{2}$ to be fixed by the boundary conditions. 
We return to the ionic densities $\rho_{\pm}$ with $X=PU$, 
\begin{subequations}\label{eq:densities}
\begin{align}
\hat{\tilde{\rho}}_{+,1}&=\nu_{1}a_{1}\sinh \mu \tilde{x} +\nu_{2}a_{2}\sinh \eta \tilde{x}\,, \\
\hat{\tilde{\rho}}_{-,1}&=a_{1}\sinh \mu \tilde{x} +a_{2}\sinh \eta \tilde{x}\,.
\end{align}
\end{subequations}
To $\mathcal{O}\left(\epsilon\right)$, \cref{eq:nofluxdimless} yields 
\begin{equation}\label{eq:nofluxbslaplace}
\partial_{\tilde{x}}\hat{\tilde{\rho}}_{\pm,1}(\pm 1,s)=-\frac{\alpha_{\pm,0}\tilde{\rho}_{\pm,0}}{s}. 
\end{equation}
Inserting \cref{eq:densities} into \cref{eq:nofluxbslaplace} yields
\begin{subequations}\label{eq:densitieslaplace}
\begin{align}
\nu_{1}a_{1}\mu\cosh \mu +\nu_{2}a_{2}\eta\cosh \eta &=-\frac{\alpha_{+}\tilde{\rho}_{+,0}}{s}\,,\\
a_{1}\mu\cosh \mu +a_{2}\eta\cosh \eta &=-\frac{\alpha_{-}\tilde{\rho}_{-,0}}{s}\,,
\end{align}
\end{subequations}
at both boundaries. 
We solve for $a_{1}$ and $a_{2}$,
\begin{subequations}\label{eq:a1a2}
\begin{align}
a_{1}&=\frac{\alpha_{-}\tilde{\rho}_{-,0}\nu_{2}-\alpha_{+}\tilde{\rho}_{+,0}}{(\nu_{1}-\nu_{2})s\,\mu\cosh \mu}\,, \\
a_{2}&=\frac{\alpha_{+}\tilde{\rho}_{+,0}-\alpha_{-}\tilde{\rho}_{-,0}\nu_{1}}{(\nu_{1}-\nu_{2})s\,\eta\cosh \eta}\,,
\end{align}
\end{subequations}
and reinsert these results into \cref{eq:densities} to find 
\begin{align}
\hat{\tilde{q}}_{1}(\tilde{x},s)&=\frac{n^2 \tilde{\rho}_{+,0}z_{+}}{s \zeta (1-\chi)}\bigg[(\alpha_{+}+\alpha_{-}\chi \nu_{2})(1+\chi\nu_{1})\frac{\sinh \mu \tilde{x}}{\mu\cosh \mu} \nn
&\quad\quad-(\alpha_{+}+\alpha_{-}\chi\nu_{1})(1+\chi\nu_{2})\frac{\sinh \eta \tilde{x}}{\eta\cosh \eta}\bigg]\,.\label{eq:hatq}
\end{align}
Now, the following local electrostatic potential
\begin{align}\label{eq:localpotential}
\hat{\tilde{\psi}}_{1}(\tilde{x},s)&=\frac{n^4(\alpha_{+}+\alpha_{-}\chi \nu_{2})}{s\zeta (1-\chi)^2 z_{-}} \frac{1+\chi\nu_{1}}{\mu^2}\left(\frac{\sinh \mu \tilde{x}}{\mu\cosh \mu}-x\right) \nn
&\quad -\frac{n^4(\alpha_{+}+\alpha_{-}\chi\nu_{1})}{s\zeta (1-\chi)^2 z_{-}} \frac{1+\chi\nu_{2}}{\eta^2}\left(\frac{\sinh \eta \tilde{x}}{\eta\cosh \eta}-x\right)\,,
\end{align}
satisfies both \cref{eq:Poissondimless,eq:nofluxdimless}. 
We use $\hat{V}_{T}(s)=\hat{\psi}(1,s)-\hat{\psi}(-1,s)$ and write $\hat{\tilde{V}}_{T}(s)\equiv \hat{\tilde{V}}_{T}^{a}(s)+\hat{\tilde{V}}_{T}^{b}(s)+\mathcal{O}\left(\epsilon^2\right)$ to find
\begin{subequations}\label{eq:thermovoltagelaplace}
\begin{align}
\hat{\tilde{V}}_{T}^{a}(s)&=\frac{2n^4\epsilon (\alpha_{+}+\alpha_{-}\chi \nu_{2}) }{s\zeta (1-\chi)^2 z_{-}}\frac{1+\chi\nu_{1}}{\mu^2}\left(\frac{\tanh \mu}{\mu}-1\right)\label{eq:thermovoltagelaplacea}\,,\\
\hat{\tilde{V}}_{T}^{b}(s)&=\frac{2n^4 \epsilon(\alpha_{+}+\alpha_{-}\chi\nu_{1})}{s\zeta (1-\chi)^2 z_{-}}\frac{1+\chi\nu_{2}}{\eta^2}\left(1-\frac{\tanh \eta}{\eta}\right)\label{eq:thermovoltagelaplaceb}\,.
\end{align}
\end{subequations}

\subsection{Solution for the Seebeck coefficient in the $t$ domain}\label{sec:theorytdomain}
The poles of $\hat{V}_{T}(s)$ at $s=0$ determine the steady state of $V_{T}(t)$. 
At $s=0$, we find $\zeta=n^2, \mu^2 =0, \eta^2 =n^2, \nu_{1} =-1/\chi$, and $ \nu_{2} =-1$.
Inserting those expressions, we find $\hat{\tilde{V}}_{T}^{a}(s)=0$ and 
\begin{equation}\label{VTs0}
\hat{\tilde{V}}_{T}^{b}(s)=-\frac{2\epsilon}{s}\frac{\alpha_{+}-\alpha_{-}}{z_{+}-z_{-}}\left(1-\frac{\tanh n}{n}\right)\,.
\end{equation}
The inverse Laplace transformation $\mathcal{L}^{-1}\{\hat{\tilde{V}}_{T}^{b}(s)\}$ now yields \cref{eq:Seebecklate} of the main text for $n\gg1$, which is a relevant simplification for us as $n>3.5\times10^6$ in our experiments.

The other poles of $\hat{\tilde{V}}_{T}(s)$ with nonzero residues appear in the $\tanh \mu$ and $\tanh \eta$ terms of \cref{eq:thermovoltagelaplace} and lie at $\mu =\pm i \mathcal{N}_{j}$ and $\eta=\pm i \mathcal{N}_{j}$, with $\mathcal{N}_{j}=(j-1/2)\pi$.
With \cref{eq:mu,eq:zeta} we write $\mu =\pm i \mathcal{N}_{j} $ to
\begin{align}
&2\mathcal{N}_{j}^2 + n^{2}+s(1+\xi)=\nn
&\hspace{1cm}=\sqrt{n^4+2n^2 s(1-\xi)\frac{\chi+1}{\chi-1}+s^2(1-\xi)^2}\,,
\end{align}
which has two solution, $s^{j}_{-}$ and $s^{j}_{+}$, for each $j$. For the experimentally relevant $n\gg1$ case, we find
\begin{subequations}
\begin{align}
s^{j}_{-} \overset{n\gg1}=&-\mathcal{N}_{j}^2\frac{1-\chi}{1-\xi\chi}+\mathcal{O}(n^{-2})\,,\\
s^{j}_{+}\overset{n\gg1}=&-\frac{ n^{2} }{ \xi } \frac{1-\xi\chi}{1-\chi}-\frac{\mathcal{N}_{j}^{2} }{\xi} \frac{1-\xi^2\chi}{1-\xi\chi}+\mathcal{O}(n^{-2})\,.
\end{align}
\end{subequations}
(The zeros of the $\tanh \eta$ term turn out to yield same $s^{j}_{\pm}$.)
For $\xi=1$ and $\chi=-1$, we find $s^{j}_{+}=-n^2-\mathcal{N}_{j}^{2}$, $s^{j}_{-}=-\mathcal{N}_{j}^{2}$, and $\nu_{1}=1$ hence $\hat{\tilde{V}}_{T}^{a}(s)=0$. 
As $V_{T}(t)$ relaxes with the Debye time for $\xi=1$ \cite{Janssen2019}, we conclude that the $s^{j}_{+}$ and $s^{j}_{-}$ solutions are associated to $\hat{\tilde{V}}_{T}^{b}(s)$ and $\hat{\tilde{V}}_{T}^{a}(s)$, respectively. 

The $n\gg1$ behavior of \cref{eq:nurandm,eq:zeta} at $s^{j}_{-}$ reads
\begin{subequations}\label{eq:nuandrsms_m}
\begin{align} 
\zeta(s^{j}_{-})&=n^2+s^{j}_{-}(1-\xi) \frac{\chi+1}{\chi-1}+\mathcal{O}\left(n^{-2}\right) \,,\\
\nu_{1} &=-\frac{1}{\chi}+s^{j}_{-}\frac{1-\xi}{\chi n^2}+\mathcal{O}\left(n^{-4}\right)\,,\\
\nu_{2} &=-1+\mathcal{O}\left(n^{-2}\right)\,,\\
\mu^2 &=-\mathcal{N}_{j}^2 +\mathcal{O}\left(n^{-2}\right)\,, 
\end{align}
\end{subequations}
where we already evaluated $\mu^2$ at $s^{j}_{-}$ and where all presented orders of $n$ were chosen with foreknowledge of the first surviving terms in the calculation below [cf. \cref{eq:thermovoltagelaplace2}].

Likewise, the $n\gg1$ behavior of \cref{eq:nurandm,eq:zeta} at $s^{j}_{+}$ reads
\begin{subequations}\label{eq:nuandrsms_p}
\begin{align} 
\zeta(s^{j}_{+})& = \frac{n^2}{\xi}\frac{1-\xi^2\chi}{1-\chi}+\frac{\mathcal{N}_{j}^{2} (1-\xi)}{ \xi}\frac{1+\xi^2\chi}{1-\xi\chi}+\mathcal{O}(n^{-2})\,,\\
\nu_{1} &=-\frac{1}{\chi\xi}+\mathcal{O}(n^{-2})\,,\\
\nu_{2}&=-\xi+\mathcal{O}(n^{-2})\,,\\
\eta&=-\mathcal{N}_{j}^{2}+\mathcal{O}(n^{-2})\,.
\end{align}
\end{subequations}

For the residues of $\hat{\tilde{V}}_{T}(s)$ at $s^{j}_{\pm}$, we inspect the terms $\tanh(\mu)/\mu-1$ and $\tanh(\eta)/\eta-1$ in \cref{eq:thermovoltagelaplace} and note that the residues of 1 at $s^{j}_{\pm}$ are zero.
For the $\tanh$ terms, we find
\begin{subequations}
\begin{align}
\frac{\tanh \mu}{\mu}\overset{\,\,s\to s^{j}_{-}}{=} &\,\,\frac{1-\chi}{1-\xi \chi}\frac{2}{s-s^{j}_{-}}+\mathcal{O}(n^{-2})\label{eq:coshclosetosm1}\,,\\
\frac{\tanh \eta}{\eta} \overset{\,\,s\to s^{j}_{+}}{=} & \,\,\frac{1-\xi^2\chi}{\xi-\xi^2\chi}\frac{2}{ s-s^{j}_{+}}+\mathcal{O}(n^{-2})\label{eq:coshclosetosm2}\,.
\end{align}
\end{subequations}
Inserting \cref{eq:nuandrsms_m,eq:coshclosetosm1} into \cref{eq:thermovoltagelaplacea} gives 
\begin{equation}\label{eq:thermovoltagelaplace2}
\hat{\tilde{V}}_{T}^{a}(s) \overset{s\to s^{j}_{-}}{\sim} \frac{\alpha_{+}z_{-}-\alpha_{-}z_{+}}{ z_{+}-z_{-}}\frac{\xi-1}{\xi z_{+}-z_{-}}\frac{4\epsilon}{\mathcal{N}_{j}^2(s-s^{j}_{-})} +\mathcal{O}(n^{-2})\,.
\end{equation}
Likewise, inserting \cref{eq:nuandrsms_p,eq:coshclosetosm2} into \cref{eq:thermovoltagelaplaceb} gives 
\begin{equation}\label{eq:thermovoltagelaplace3}
\hat{\tilde{V}}_{T}^{b}(s)\overset{s\to s^{j}_{+}}{\sim}4\epsilon\frac{\xi\alpha_{+}-\alpha_{-}}{\xi z_{+}-z_{-}}\frac{1}{\mathcal{N}_{j}^{2}( s-s^{j}_{+})}+\mathcal{O}(n^{-2})\,.
\end{equation}
Calculating $V_{T}(t)=\sum_{\{0,s^{j}_{-}, s^{j}_{+}\}}\textrm{Res}\left(\hat{V}_{T}(s)\exp{(s\tilde{t})},s\right)$ now gives 
\begin{subequations}\label{eq:VTfinal}
\begin{equation}
V_{T}(t)=V_{T, \mathrm{late}}+V_{T, \mathrm{dif}}(t)+V_{T, D}(t)+\mathcal{O}\left(n^{-1}, \epsilon\right)\,,
\end{equation}
with
\begin{align}
\frac{V_{T, \mathrm{late}}}{\Delta T}&=-\frac{2\kb }{e} \frac{\alpha_{+} -\alpha_{-}}{z_{+}-z_{-}}\,,\label{eq:Vlate}\\
\frac{V_{T, \mathrm{dif}}(t)}{\Delta T}&=\frac{4\kb }{e}\frac{D_{+}-D_{-}}{D_{+} z_{+}-D_{-}z_{-}}\frac{\alpha_{+}z_{-}-\alpha_{-}z_{+}}{ z_{+}-z_{-}} \nn
&\quad\times\sum_{j=1}^{\infty}\frac{\exp{\left[-t \mathcal{N}_{j}^2 D_{a}/L^2 \right]}}{\mathcal{N}_{j}^2}\,,\label{eq:VTdif}\\
\frac{V_{T, D}(t)}{\Delta T}&=\frac{2\kb }{e} \frac{D_{+}\alpha_{+}-D_{-}\alpha_{-}}{D_{+} z_{+}-D_{-}z_{-}}\exp{\left[-tD_{m}/\lambda^2\right]}\,,\label{eq:VTD}
\end{align}
\end{subequations}
where we used $\sum_{j=1}^{\infty}\mathcal{N}_{j}^{-2}=1/2$ for the sum over $s^{j}_{+}$, leading to $V_{T, D}(t)$, and where $D_{m}$ and $D_{a}$ are valency-weighted arithmetic and harmonic means of the cationic and anionic diffusivities
\begin{equation}\label{eq:amphotericdiffusivity2}
D_{m}=\frac{D_{+}z_{+}-D_{-}z_{-}}{z_{+}-z_{-}}\,,\qquad 
D_{a}=\frac{(z_{+}-z_{-})D_{+}D_{-}}{z_{+}D_{+}-z_{-}D_{-}}\,.
\end{equation}
For $z_{+}=-z_{-}=1$, \cref{eq:VTfinal} correctly reduces to Eq.~(16) of \cit{Janssen2019} (up to the different minus-sign convention for $V_{T}$).

From \cref{eq:VTfinal} we find the transient Seebeck coefficient $S(t) =-V_{T}(t)/\Delta T$ as
\begin{align}\label{eq:Sttheory0}
S(t) &=S_\mathrm{late}+2(S_\mathrm{early}-S_\mathrm{late})\sum_{j=1}^{\infty}\frac{\exp{\left[- \mathcal{N}_{j}^2 t /(\mathcal{N}_{1}^2\tau_\mathrm{dif}) \right]}}{\mathcal{N}_{j}^2}\nn
&\quad-S_\mathrm{early}\exp{\left[-t/\tau_{D}\right]}+\mathcal{O}\left(n^{-1}, \epsilon\right)\,,
\end{align}
with $S_\mathrm{late}$ and $S_\mathrm{early}$ from \cref{eq:Seebecklate,eq:Seebeckearly} of the main text and with $\tau_\mathrm{dif}= 4L^2/(\pi^2 D_{a})$ the diffusion time and $\tau_{D}=\lambda^2/D_{m}$ the Debye time---generalizing our definitions of \cref{sec:Introduction} of the main text, where we considered $D_{+}=D_{-}$.

\begin{figure}
\includegraphics[width=8.6cm]{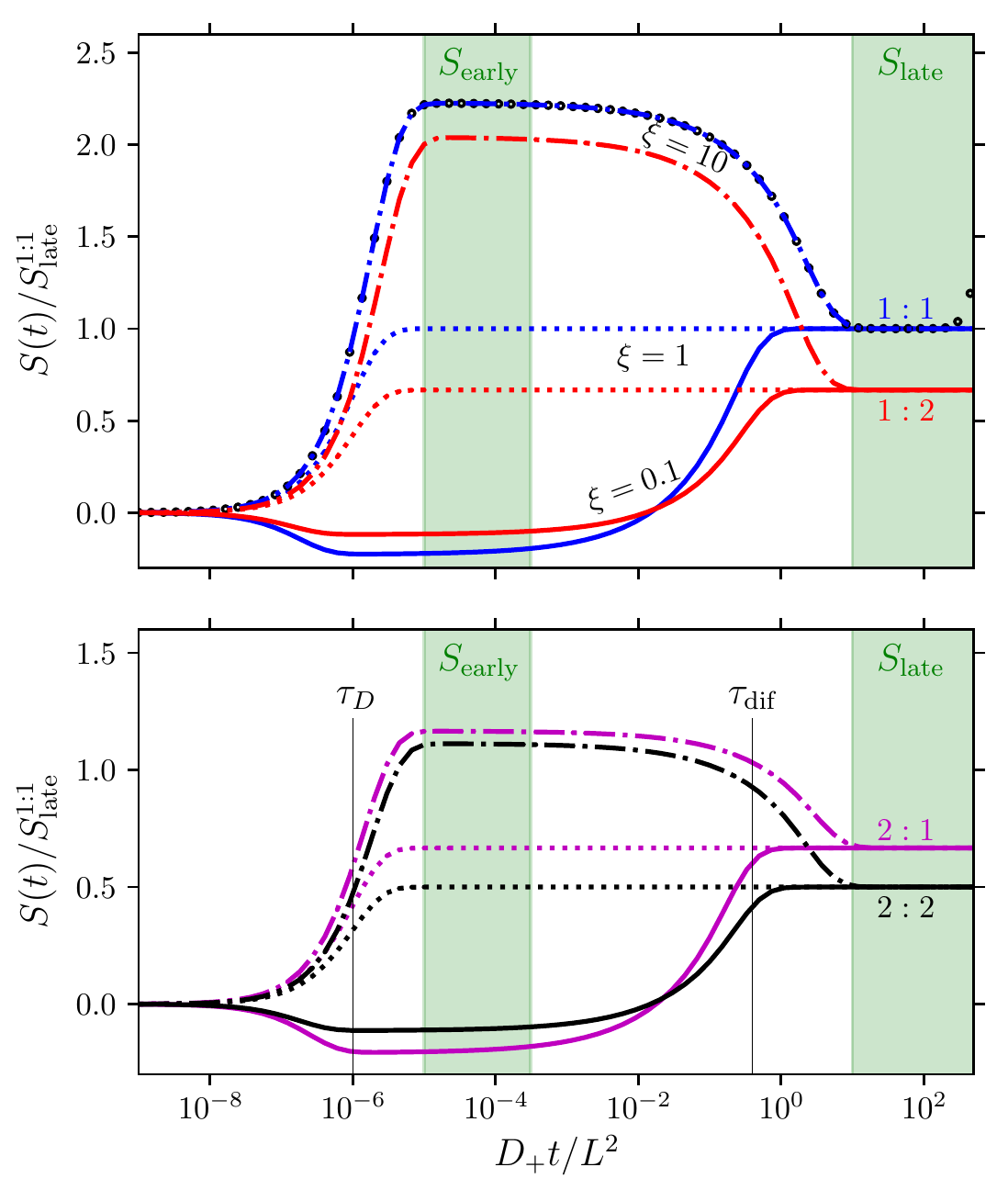}
\caption{
$S(t)$ [\cref{eq:Sttheory0}] normalized to $S_\mathrm{late}$ [\cref{eq:Seebecklate} of the main text] of a $1:1$ electrolyte, which we denote $S_\mathrm{late}^{1:1}$. 
We show several valencies [$1:1$ (blue), $1:2$ (red), $2:1$ (magenta), $2:2$ (black)] and $\xi\equiv D_{+}/D_{-}=0.1, 1, 10$ (solid, dotted, and dash-dotted lines), and use $\alpha_{+}=0.5$, $\alpha_{-}=0.1$, $n\equiv L/\lambda=1000$, and $\max(j)=500$ throughout. 
The shaded areas correspond to times for which $S(t)$ takes its $S_\mathrm{early}$ and \mbox{$S_\mathrm{late}$ values}. 
We also indicate the relaxation times $\tau_{D}$ and $\tau_\textrm{dif}$ for $\xi=1$ ($1/n^2$ and $4/\pi^2$, respectively, in units of $L^2/D_{+}$); these times are slightly shifted for other $\xi$ as $D_{a}$ and $D_{m}$ [\cref{eq:amphotericdiffusivity2}] are $\xi$-dependent. 
For the $1:1$ electrolyte with $\xi=10$, we show with black circles the result of substituting \cref{eq:jaegerandcarslaw} into \cref{eq:Sttheory0}, using $\max(k)=10$.}
\label{figS1}
\end{figure}

At $t=0$, the time-dependent exponents in \cref{eq:Sttheory0} are unity. 
Again using $\sum_{j=1}^{\infty}\mathcal{N}_{j}^{-2}=1/2$ then yields $S(0)=0$.
Then, the early Debye-time relaxation of $S(t)$ comes from the last term in \cref{eq:Sttheory0}.
After this relaxation, for times $t\gg \tau_{D}\land t \ll \tau_\mathrm{dif}$, we have $\exp{\left[-t/\tau_{D}\right]}\approx 0$ and $\exp{\left[- (2j-1)^{2}t /\tau_\mathrm{dif} \right]}\approx 1$ and, using the same sum identity, we find $S(t)=S_\mathrm{early}$.
Finally, at late times $t \gg \tau_\mathrm{dif}$, $S(t)$ relaxes to $S_{\mathrm{late}}$.
This $S(t)$ behavior is also visible in \cref{figS1}, where we plot \cref{eq:Sttheory0} for several valencies and $\xi=D_{+}/D_{-}$ at fixed $\alpha_{+}=0.5$, $\alpha=0.1$, $n=10^{3}$, and $\max(j)=500$. 
We also see there that $S_\mathrm{early}=S_\mathrm{late}$ for $\xi=1$, that higher valencies generally lead to smaller $S(t)$, and that higher $\xi$ lead to slower relaxation (for these $\alpha_{\pm}$).
Moreover, while \cref{eq:VTfinal,eq:Sttheory0} are invariant under $+\leftrightarrow-$, \cref{figS1} shows that $S(t)$ is not invariant under $\xi\to1/\xi$ at fixed $\alpha_{\pm}$. 

Interestingly, the sum in \cref{eq:Sttheory0} can be rewritten with an identity that is implicit in Eqs.~(10) and (11) on page 97 of \cit{carslaw1959conduction} (up to a factor-$\pi$ typo in their exponent),
\begin{equation}\label{eq:jaegerandcarslaw}
\sum_{j=1}^{\infty}\frac{\exp{\left(- \mathcal{N}_{j}^{2}\theta \right)}}{\mathcal{N}_{j}^{2}}=\frac{1}{2}-\sqrt{\frac{\theta}{\pi}}-2\sqrt{\theta}\sum_{k=1}^{\infty}(-1)^{k}\mathrm{ierfc}\left(\frac{k}{\sqrt{\theta}}\right)\,.
\end{equation}
\Cref{eq:jaegerandcarslaw} is useful as the sum on its left-hand side converges quickly for $\theta\gtrapprox 1$, while, conversely, the sum on its right-hand side does so for $\theta\ll1$.
Accordinlgy, for the $\xi=10$, $1:1$ electrolyte we substitute \cref{eq:jaegerandcarslaw} into \cref{eq:Sttheory0} and show $S(t)$ with black circles.
Truncating the error function expression already at $k=10$, this expression approximates $S(t)$ very well until $D_{-}t/L^2=10^2$. 
For larger $\max(k)$, the correspondence becomes even better (not shown).

\subsection{Seebeck coefficient for a slowly applied $\Delta T$}\label{sec:slowDeltaT}
In the experiments of \cref{sec:experiments} of the main text, the temperature difference increases roughly as $\Delta T(t)=\Delta T_{\infty} \left[1-\exp (-t/\tau_\mathrm{ap})\right]$, with $\Delta T_{\infty}$ its late-time asymptote and with $\tau_\mathrm{ap}$ a characteristic timescale. 
We replace our assumption of an instantaneous steady-state temperature profile $T(x)$ by $T(x,t)=T_{0}+\Delta T_{\infty} \left[1-\exp (-t/\tau_\mathrm{ap})\right] (x+L)/(2L)$.
Redefining $\epsilon = \Delta T_{\infty}/T_{0}$ and tracing our steps of \cref{sec:dimless,sec:solinsdomain,sec:theorytdomain}, we see that \cref{eq:nofluxdimless} changes to 
\begin{equation}
\tilde{J}_{\pm}(\pm 1,\tilde{t}\,)=\partial_{\tilde{x}} \tilde{\rho}_{\pm}+\frac{\alpha_{\pm}\tilde{\rho}_{\pm} \left[1-\exp (-\tilde{t}/\tilde{\tau}_\mathrm{ap})\right] }{\tilde{T}}=0\,,
\end{equation}
with $\tilde{\tau}_\mathrm{ap}= \tau_\mathrm{ap} D_{+}/L^2$, and that \cref{eq:nofluxbslaplace} changes to 
\begin{equation}
\partial_{\tilde{x}}\hat{\tilde{\rho}}_{\pm,1}(\pm 1,s)=-\frac{\alpha_{\pm,0}\tilde{\rho}_{\pm,0}}{s(1+\tilde{\tau}_\mathrm{ap} s)}. 
\end{equation}
The same factor $1/(1+\tilde{\tau}_\mathrm{ap} s)$ then enters \cref{eq:densitieslaplace,eq:a1a2,eq:hatq,eq:localpotential,eq:thermovoltagelaplace,VTs0,eq:thermovoltagelaplace2,eq:thermovoltagelaplace3}. 
In particular, \cref{eq:thermovoltagelaplace} now reads
\begin{subequations}\label{eq:VTlaplacetau2}
\begin{align}
\hat{\tilde{V}}_{T}^{a}(s)&=\frac{2n^4\epsilon (\alpha_{+}+\alpha_{-}\chi \nu_{2}) }{s(1+\tilde{\tau}_\mathrm{ap} s)\zeta (1-\chi)^2 z_{-}}\frac{1+\chi\nu_{1}}{\mu^2}\left(\frac{\tanh \mu}{\mu}-1\right)\label{eq:thermovoltagelaplacea2}\,,\\
\hat{\tilde{V}}_{T}^{b}(s)&=\frac{2n^4 \epsilon(\alpha_{+}+\alpha_{-}\chi\nu_{1})}{s(1+\tilde{\tau}_\mathrm{ap} s)\zeta (1-\chi)^2 z_{-}}\frac{1+\chi\nu_{2}}{\eta^2}\left(1-\frac{\tanh \eta}{\eta}\right)\label{eq:thermovoltagelaplaceb2}\,.
\end{align}
\end{subequations}
Clearly, at $s=0$, the factor $1/(1+\tilde{\tau}_\mathrm{ap} s)$ is unity; hence, $V_{T, \mathrm{late}}$ is unaffected.
Evaluating the poles $s^{j}_{\pm}$ now yields
\begin{subequations}\label{eq:Vtauapp}
\begin{align}
\frac{V_{T, \mathrm{dif}}(t)}{\Delta T_{\infty}}&=\frac{4\kb }{e}\frac{D_{+}-D_{-}}{D_{+} z_{+}-D_{-}z_{-}}\frac{\alpha_{+}z_{-}-\alpha_{-}z_{+}}{ z_{+}-z_{-}} \nn
&\quad\times\sum_{j=1}^{\infty}\frac{\exp{\left[- \mathcal{N}_{j}^2 t /(\mathcal{N}_{1}^2\tau_\mathrm{dif}) \right]}}{\mathcal{N}_{j}^2[1- \mathcal{N}_{j}^2\tau_\mathrm{ap} /( \mathcal{N}_{1}^2\tau_\mathrm{dif} )]}\label{eq:VTtaudif}\,,\\
\frac{V_{T, D}(t)}{\Delta T_{\infty}}&=\frac{2\kb }{e} \frac{D_{+}\alpha_{+}-D_{-}\alpha_{-}}{D_{+} z_{+}-D_{-}z_{-}}\frac{\exp{\left[-t/\tau_{D}\right]}}{1-\tau_\mathrm{ap}/\tau_{D}}\label{eq:VTtauD}\,.
\end{align}
\end{subequations}
Next to these modifications of \cref{eq:VTdif,eq:VTD}, $V_{T}(t)$ gets new terms from the pole $s=-1/\tilde{\tau}_\mathrm{ap}$ of \cref{eq:VTlaplacetau2}.
In our experiments, $\tau_\mathrm{ap}\sim\SI{e2}{\second}$ and $D_{+}/L^2\sim \SI{e-3}{\per\second}$; hence, $\tilde{\tau}_\mathrm{ap}\sim 10^{-1}$, which is 12 order of magnitude larger than $1/n^{2}\sim 10^{-13}$. 
Expanding $\zeta$ [\cref{eq:zeta}] for large $n$, we thus assume that $\tilde{\tau}_\mathrm{ap}\gg 1/n^{2}$ and find
\begin{equation}\label{eq:zetaexpansion}
\zeta=n^2-\frac{1-\xi}{\tilde{\tau}_\mathrm{ap}} \frac{\chi+1}{ \chi-1} +\mathcal{O}(n^{-2}) \,.
\end{equation}
At $s=-1/\tilde{\tau}_\mathrm{ap}$, the terms $\mu, \eta, \nu_{1}$, and $\nu_{2}$ in \cref{eq:VTlaplacetau2} read
\begin{subequations}
\begin{align}
\mu^2 &=\frac{1}{\tilde{\tau}_\mathrm{ap}}\frac{1-\chi\xi}{\chi-1}+\mathcal{O}(n^{-2}) \,,\\
\eta^2 &=n^2+\frac{1}{\tilde{\tau}_\mathrm{ap}}\frac{ \xi-\chi}{\chi-1} +\mathcal{O}(n^{-2}) \,,\\
\nu_{1}&=- \frac{1}{\chi}+\frac{\xi-1}{n^{2} \tilde{\tau}_\mathrm{ap} \chi }+\mathcal{O}(n^{-4}) \,,\\
\nu_{2}&=-1+\frac{\xi-1}{n^{2} \tilde{\tau}_\mathrm{ap} }+\mathcal{O}(n^{-4}) \,.
\end{align}
\end{subequations}
Calculating the residue of \cref{eq:thermovoltagelaplacea2} at $-1/\tilde{\tau}_\mathrm{ap}$ now gives 
\begin{align}\label{eq:tauappV1}
&\textrm{Res}\left(\hat{V}^{a}_{T}(s)\exp{(s\tilde{t})},s=-1/\tilde{\tau}_\mathrm{ap}\right)=\nn
&=-\frac{4\kb \Delta T_{\infty}}{e} \frac{ \alpha_{+}z_{-}-\alpha_{-}z_{+} }{z_{+}-z_{-} }\frac{D_{+}-D_{-}}{D_{+}z_{+}-z_{-}D_{-} }\nn
&\quad\quad \times \sum_{j=1}^{\infty}\frac{\exp{\left(-t/\tau_\mathrm{ap} \right)}}{\mathcal{N}_{j}^{2}[1- \mathcal{N}_{j}^2\tau_\mathrm{ap} /( \mathcal{N}_{1}^2\tau_\mathrm{dif} )]}+\mathcal{O}(n^{-2})\,,
\end{align}
where we used $1-\tanh \mu /\mu = 2\sum_{j=1}^{\infty}1/[\mathcal{N}_{j}^{2}(1+\mathcal{N}_{j}^{2}/\mu^2)]$. 
Likewise, for \cref{eq:thermovoltagelaplaceb2} we find
\begin{align}\label{eq:tauappV2}
&\textrm{Res}\left(\hat{V}^{b}_{T}(s)\exp{(s\tilde{t})},s=-1/\tilde{\tau}_\mathrm{ap}\right)=\nn
&\quad\quad=\frac{2\kb \Delta T_{\infty}}{e}\frac{\alpha_{+}- \alpha_{-}}{z_{+}-z_{-} }\exp\left(-t/\tau_\mathrm{ap}\right)-O(n^{-2})\,.
\end{align}
Combining \cref{eq:Vlate,eq:Vtauapp,eq:tauappV1,eq:tauappV2} gives 
\begin{widetext}
\begin{equation}\label{eq:Vttheory}
\frac{V_{T}(t)}{\Delta T_{\infty}}= -S_\mathrm{late}\left[1-\exp\left(-\frac{t}{\tau_\mathrm{ap}}\right)\right]+\frac{8(S_\mathrm{late}-S_\mathrm{early})}{\pi^2}
\sum_{j=1}^{\infty}\frac{\exp{\left[- (2j-1)^{2}t /\tau_\mathrm{dif} \right]}-\exp{\left(-t/\tau_\mathrm{ap} \right)}}{ (2j-1)^{2}[1- (2j-1)^{2} \tau_\mathrm{ap} /\tau_\mathrm{dif} ]}+\mathcal{O}\left(n^{-1}, \epsilon\right)\,,
\end{equation}
\end{widetext}

\begin{figure}[h]
\includegraphics[width=8.6cm]{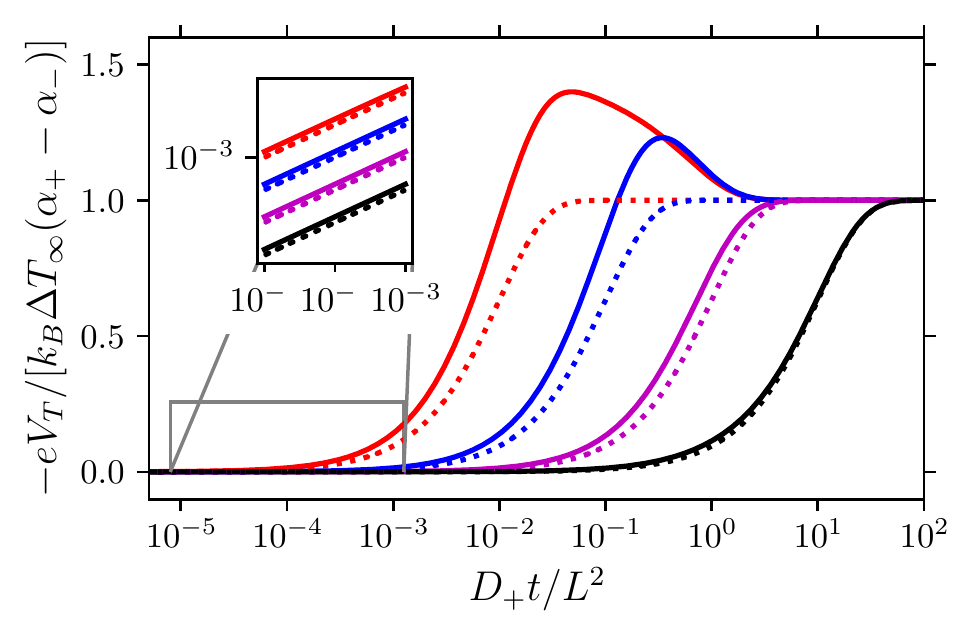}
\caption{
$V_{T}$ [\cref{eq:Vttheory}] (lines) and the first term of the right hand side of \cref{eq:Vttheory} (dotted) 
for several $\tau_\mathrm{ap}D_{+}/L^2$ [$0.01$ (red), $0.1$ (blue), $1$ (magenta), $10$ (black)] with $\xi\equiv D_{+}/D_{-}=2$, $\alpha_{+}=0.5$, $\alpha_{-}=0.1$, and $\max(j)=500$ throughout. }
\label{figS2}
\end{figure}
Note that \cref{eq:VTtauD} does not contribute at this order in $n$, as $V_{T, D}(t)\sim 1/n^2$ when $\tilde{\tau}_\mathrm{ap}\gg 1/n^{2}$.
The main result \cref{eq:Sttheory} of the main text now follows from dividing $V_{T}(t)$ [\cref{eq:Vttheory}] by $\Delta T(t)=\Delta T_{\infty}\left[1-\exp\left(-t/\tau_\mathrm{ap}\right)\right]$.
Note also that \cref{fig:VTtau} of the main text differs by the same factor from \cref{figS2}, described below. 

\Cref{figS2} shows \cref{eq:Vttheory} (lines) and the first term of the right hand side of \cref{eq:Vttheory} (dotted lines).
For $\tau_\mathrm{ap}>\tau_\textrm{dif}$ ($\tau_\textrm{dif}D_{+}/L^2=0.61$ for the considered parameters), the difference between solid and dotted lines is minor, meaning that $V_{T}(t)\approx-S_\mathrm{late}\Delta T(t)$.
Conversely, the contribution of the second term on the right hand side \cref{eq:Vttheory} is apparent for smaller $\tau_\mathrm{ap}$.
At early times, scrutinised in the inset, $V_{T}(t)=-S_\mathrm{late}\Delta T(t)$ systematically underestimates $V_{T}(t)$ by a factor 1.5, which coincides with the ratio $S_\mathrm{early}/S_\mathrm{late}$ for the used parameters [cf.~\cref{fig:VTtau} of the main text]. 
Hence, at early times, $V_{T}(t)\approx-S_\mathrm{early}\Delta T(t)$.
Indeed, in \cref{fig:VTtau} of the main text we see that $S(t\ll\tau_\textrm{dif})=S_\mathrm{early}$, independent of $\tau_\textrm{ap}/\tau_\textrm{dif}$.

\section{Shortcut to $S_\mathrm{late}$}\label{sec:shortcut}
As is clear from \cref{eq:Vttheory}, the steady state of \cref{eq:1dflux,eq:electrokinetic,eq:initandbc} of the main text is characterised by the Seebeck coefficient $S_\mathrm{late}$.
This expression, however, can be derived much quicker from the same set of equations.
Our derivation below is largely analogous to \cit{majee2011} who studied \mbox{$E=-\partial_x \psi$} instead of $\psi$.
With \cref{eq:1dflux} of the main text, we rewrite $z_{+}J_{+}(x)/D_{+}+z_{-}J_{-}(x)/D_{-}=0$---which obviously holds at steady state)---to
\begin{equation}
-\partial_{x} q-\frac{z_{+}^2\rho_{+}+z_{-}^2\rho_{-}}{\kb T}e\partial_{x} \psi =\frac{z_{+}\rho_{+}Q^{*}_{+}+z_{-}\rho_{-}Q^{*}_{-}}{\kb T^2}\frac{\Delta T}{2L}\,.
\end{equation}
With a small-$\epsilon$ expansion we find 
\begin{equation}
-\partial_{\tilde{x}} \tilde{q}_{1}-(z_{+}^2\tilde{\rho}_{+,0}+z_{-}^2\tilde{\rho}_{-,0}) \partial_{\tilde{x}} \tilde{\psi}_{1} =z_{+}\tilde{\rho}_{+,0}\left(\alpha_{+}-\alpha_{-}\right)\,.
\end{equation}
Inserting \cref{eq:Poissondimless_b} yields
\begin{equation}\label{eq:thirdorder}
\frac{1}{n^2}\partial_{\tilde{x}}^{3} \tilde{\psi}_{1}- \partial_{\tilde{x}} \tilde{\psi}_{1} =\frac{\alpha_{+}-\alpha_{-}}{z_{+}-z_{-}}\,,
\end{equation}
which is solved by
\begin{equation}
 \tilde{\psi}_{1}(\tilde{x})=b_{1}+b_{2}\exp({n\tilde{x}})+b_{3}\exp({-n\tilde{x}})-\tilde{x}\frac{\alpha_{+}-\alpha_{-}}{z_{+}-z_{-}}\,,
\end{equation}
wherein three constants, $b_{1}, b_{2}$, and $b_{3}$, appear. 
As $\psi_{1}$ is only defined up to a constant, we can set $b_{1}=0$ without loss of generality. 
From \cref{eq:continuity_ions,eq:initandbcb,eq:initandbcd} of the main text follows that the initially charge-neutral electrolyte stays globally charge neutral at later times as well: $\int_{-L}^{L} \dif x\, q(x)=0$.
Inserting \cref{eq:Poisson} of the main text we find $\partial_{\tilde{x}}\tilde{\psi}_{1}(1)-\partial_{\tilde{x}}\tilde{\psi}_{1}(-1)=0$, which fixes $b_{3}=-b_{2}$. 
We use \cref{eq:initandbcc} of the main text to fix the remaining constant to $b_{2}= (\alpha_{+}-\alpha_{-})/[2(z_{+}-z_{-})n\cosh n]$. 
The resulting electrostatic potential
\begin{equation}\label{eq:psi_sol}
 \tilde{\psi}_{1}(\tilde{x})=\frac{\alpha_{+}-\alpha_{-}}{z_{+}-z_{-}} \left(\frac{\sinh n\tilde{x}}{n \cosh n}-\tilde{x}\right)\,.
\end{equation}
yields 
\begin{equation}
S_\mathrm{late}=2\frac{\alpha_{+}-\alpha_{-}}{z_{+}-z_{-}} \left(1- \frac{\tanh n}{n }\right)\,,
\end{equation}
which is equivalent to \cref{VTs0} and which reduces to \cref{eq:Seebecklate} of the main text for $n\gg1$. 
For a nonzero surface charge $\sigma$, the additional $\sigma$-dependent term also drops out for $n\gg1$, again yielding \cref{eq:Seebecklate}.

Reinserting \cref{eq:psi_sol} into \cref{eq:thirdorder}, we see that the diffusion term of \cref{eq:thirdorder},
\begin{equation}
\frac{1}{n^1}\partial_{\tilde{x}}^{3} \tilde{\psi}_{1}=\frac{\alpha_{+}-\alpha_{-}}{z_{+}-z_{-}} \frac{\cosh n\tilde{x}}{\cosh n}\,,
\end{equation}
 is \textit{not} small compared to the electromigration term, 
\begin{equation} 
\partial_{\tilde{x}} \tilde{\psi}_{1}=\frac{\alpha_{+}-\alpha_{-}}{z_{+}-z_{-}} \left(\frac{\cosh n\tilde{x}}{\cosh n}-1\right)
\end{equation}
---even at $n\gg1$, they are both $\mathcal{O}(1)$. 
Rather, $\partial_{\tilde{x}}^{3} \tilde{\psi}_{1}/n^2$ exactly cancels an opposite term in $\partial_{\tilde{x}} \tilde{\psi}_{1}$.
Assuming $q(x,t)=0$ to hold everywhere does not properly account for the fact a nonzero thermovoltage is ultimately caused by the regions, however small, where $q(x,t)\neq0$.

We have not found a similar shortcut to \cref{eq:Seebeckearly} of the main text. 
Not only is the commonly employed assumption of $\partial_{x}\rho_{i}=0$ incorrect at intermediate times $t\gg \tau_{D}\land t \ll \tau_\mathrm{dif}$, so is the assumption of (any combination of) ionic currents $J_{i}(x,t)$ to vanish. 

\onecolumngrid
\clearpage

\section{Supplementary Figures}

\begin{figure}[h]
\includegraphics[width=12.5cm]{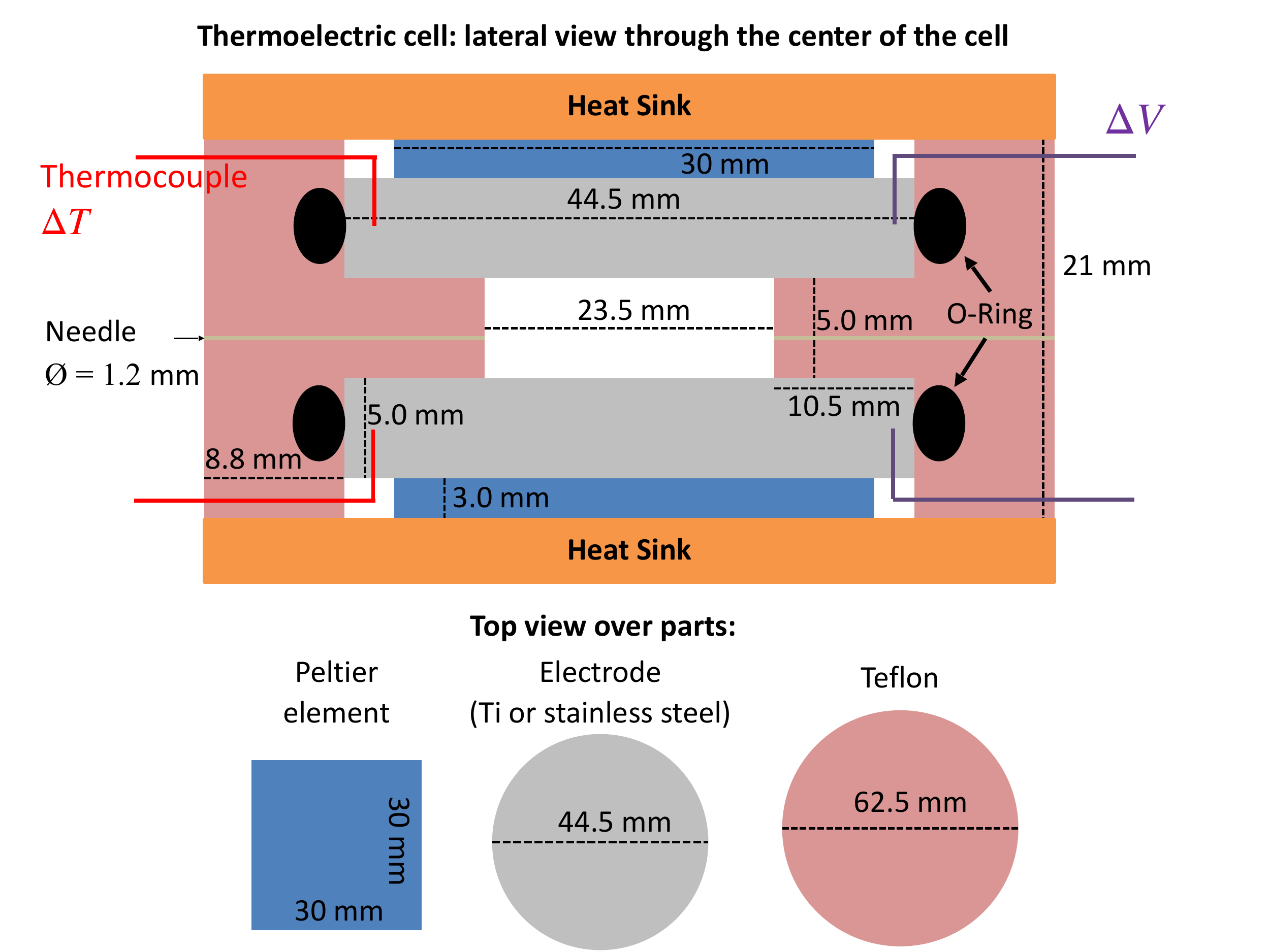}
\vspace{-0.3cm}
\caption{\textbf{Schematic of our experimental setup}}
\label{figS3}
\end{figure}

\begin{figure}[h]
\includegraphics[width=16.2cm]{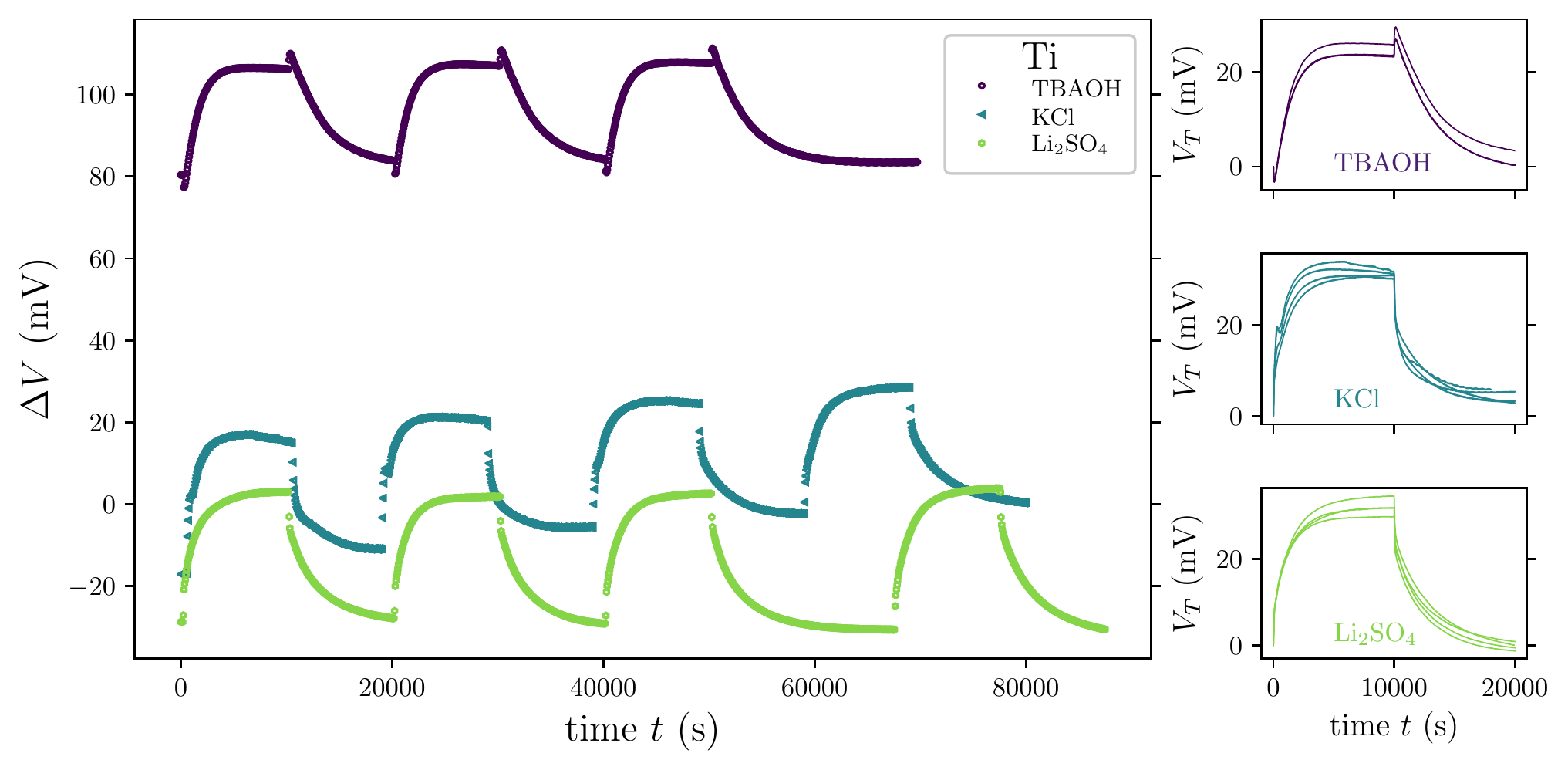}
\vspace{-0.3cm}
\caption{\textbf{Reversibility of $\Delta V(t)$ and spread in $V_{T}(t)$ between different heating-cooling cycles.} 
Several heating-cooling cycles of TBAOH, KCl, and Li$_{2}$SO$_{4}$ near Ti electrodes show the reproducibility of the thermovoltage (we show every 20th data point). 
To the right, we superimpose all these heating-cooling cycles, setting $V_{T}(t)=0$ at the start of each cycle. We see that $V_{T}(t)$ returns to 0 within a few mV after each heating-cooling cycle of TBAOH (what appears as a thicker line in the bottom actually consists of two different cycles) and Li$_{2}$SO$_{4}$. $V_{T}(t)$ data for KCl is less reversible. }
\label{figS4}
\end{figure}

\begin{figure}
\includegraphics[width=14cm]{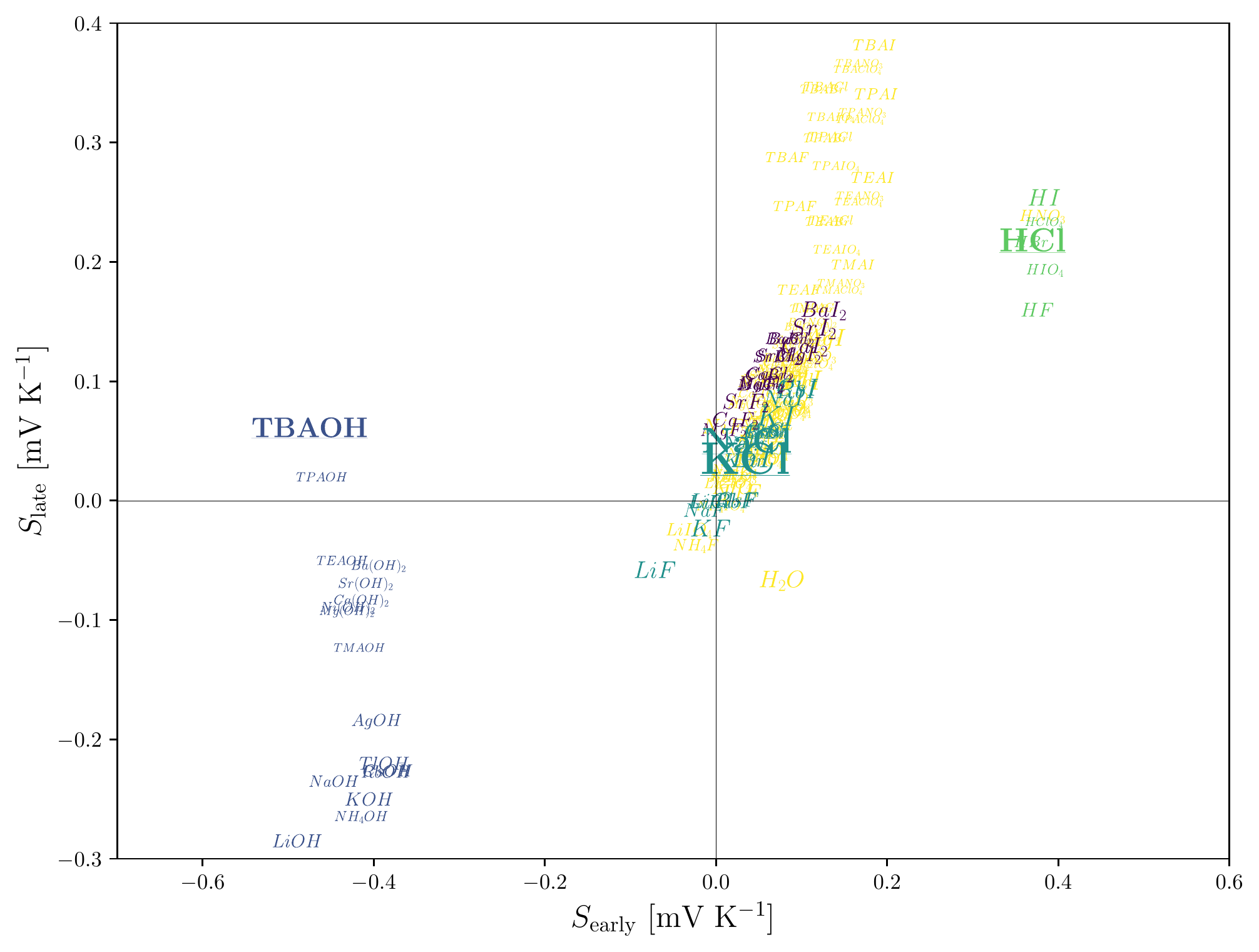}
\caption{\textbf{$S_\mathrm{early}$ vs $S_\mathrm{late}$ of all possible cation-anion combinations with the $Q_{i}^{*}$ data of \cit{agar1989}}: {\color{nr2}20 alkalihalides}, {\color{nr0}16 alkaline earth metal halides}, {\color{nr1}17 hydroxides}, {\color{nr3}7 acids}, and {\color{nr4}84 other} combinations. 
We see that these different types of electrolytes cluster in this representation. 
Moreover, TBAOH is one of the few electrolytes for which $\textrm{sgn} (S_\mathrm{early})\neq \textrm{sgn}(S_\mathrm{late})$, which we also found in our experiments. 
With underlines and boldface we highlighted TBAOH, HCl, and KCl that we studied experimentally in the main text. (We could not predict $S$ of the sulfate salts, as \cit{agar1989} does not report $\alpha_{-}$ of the sulfate anion.) 
While several of our experiments yielded $S_\mathrm{late}\approx \SI{2}{\milli\volt\per\kelvin}$, the data shown here suggests that $S_\mathrm{late}< \SI{0.4}{\milli\volt\per\kelvin}$.}
\label{figS5}
\end{figure}

\end{document}